\documentclass[prb,nobibnotes,altaffilletter,amsmath,amssymb,amsfonts]{revtex4}
\usepackage[latin1]{inputenc}
\usepackage{graphicx}
\usepackage{amsmath}
\usepackage{bm}
\usepackage{latexsym}
\usepackage{epsfig}
\newcommand{\br}{{\bf r}}
\newcommand{\bF}{{\bf F}}
\newcommand{\tr}{\tilde{\bf r}}

\newcommand{\tF}{\tilde{\bf F}}
\newcommand{\tU}{\tilde{U}}
\newcommand{\tti}{\tilde{t}}
\newcommand{\tm}{\tilde{m}}
\newcommand{\bR}{{\bf r}_1, ..., {\bf r}_N}
\newcommand{\tR}{\tilde{\bf r}_1, ..., \tilde{\bf r}_N}
\newcommand{\angleb}[1]{\langle #1 \rangle}

\newcommand{\cv}{c_V^{\rm ex}}
\newcommand{\ccv}{C_V^{\rm ex}}
\newcommand{\bv}{\beta_V^{\rm ex}}
\newcommand{\sex}{S_{\rm ex}}

\begin{document}
\title{Pressure-energy correlations in liquids. IV. ``Isomorphs'' in liquid state diagrams}

\author{Nicoletta Gnan}
\email{ngnan@ruc.dk}
\affiliation{DNRF Center ``Glass and Time'', IMFUFA, Dept. of Sciences, Roskilde University, P.O. Box 260, DK-4000 Roskilde, Denmark}
\author{Thomas B. Schr{\o}der}
\email{tbs@ruc.dk}
\affiliation{DNRF Center ``Glass and Time'', IMFUFA, Dept. of Sciences, Roskilde University, P.O. Box 260, DK-4000 Roskilde, Denmark}
\author{Ulf R. Pedersen}
\email{urp@ruc.dk}
\affiliation{DNRF Center ``Glass and Time'', IMFUFA, Dept. of Sciences, Roskilde University, P.O. Box 260, DK-4000 Roskilde, Denmark}
\author{Nicholas P. Bailey}
\email{nbailey@ruc.dk}
\affiliation{DNRF Center ``Glass and Time'', IMFUFA, Dept. of Sciences, Roskilde University, P.O. Box 260, DK-4000 Roskilde, Denmark}
\author{ Jeppe C. Dyre }
\email{dyre@ruc.dk}
\affiliation{DNRF Center ``Glass and Time'', IMFUFA, Dept. of Sciences, Roskilde University, P.O. Box 260, DK-4000 Roskilde, Denmark}

\date{\today}
\begin{abstract}
This paper is the fourth in a series devoted to identifying and explaining the properties of strongly correlating liquids, i.e., liquids where virial and potential energy correlate better than 90\% in their thermal equilibrium fluctuations in the NVT ensemble. For such liquids we here introduce the concept of ``isomorphic'' curves in the state diagram. A number of thermodynamic, static, and dynamic isomorph invariants are identified. These include the excess entropy, the isochoric specific heat, reduced-unit static and dynamic correlation functions, as well as reduced-unit transport coefficients. The dynamic invariants apply for both Newtonian and Brownian dynamics. It is shown that after a jump between isomorphic state points the system is instantaneously in thermal equilibrium; consequences of this for generic aging experiments are discussed. Selected isomorph predictions are validated by computer simulations of the Kob-Andersen binary Lennard-Jones mixture, which is a strongly correlating liquid. The final section of the paper relates the isomorph concept to phenomenological melting rules, Rosenfeld's excess entropy scaling, Young and Andersen's approximate scaling principle, and the two-order parameter maps of Debenedetti and coworkers. This section also shows how the existence of isomorphs implies an ``isomorph filter'' for theories for the non-Arrhenius temperature dependence of viscous liquids' relaxation time, as well as explains isochronal superposition for strongly correlating viscous liquids.
\end{abstract}

\maketitle

\section{Introduction}

How much does knowledge of a system's thermal equilibrium fluctuations at one state point tell us about its behavior at other state points? In principle, complete knowledge of the fluctuations provides enough information to determine the density of states, from which the free energy at other state points may be calculated. In practice, only second-order moments of the fluctuations may be determined reliably. These generally give little knowledge of the system's properties away from the state point in question. It was recently shown that a large class of liquids exhibit strong correlations between their virial and potential energy (NVT) thermal equilibrium fluctuations.\cite{ped08a,ped08b,I,II,III,cos08} Such liquids have a hidden (approximate) scale invariance.\cite{III,sch08b} Because of this, important global information about the system may be obtained from knowledge of the virial and potential energy fluctuations' second-order moments at one state point. This unusual situation in statistical mechanics is the background of the present paper, which is the fourth in a series\cite{I,II,III} devoted to identifying and explaining the properties of strongly correlating liquids. 

Paper I\cite{I} of the series presented results from computer simulations of 13 different liquids. The results show that van der Waals and metallic liquids are strongly correlating, whereas hydrogen-bonding liquids like methanol and water are not. Likewise, covalent and ionic liquids are not expected to be strongly correlating because competing interactions generally spoil the correlations. Paper II\cite{II} gave a thorough analysis of the cause of the strong correlations, which is briefly recapitulated below. It was shown how to qualify the simple explanation of the correlations given in our first publication (Ref. \onlinecite{ped08a}), where strong correlations were argued to derive from particle-particle close encounters, thus probing only the repulsive part of the potential which is in many cases well approximated by an inverse power law. This explanation must be qualified in order to explain the occurrence of strong correlations at low temperatures and/or low pressures, as well as in the crystalline state. A number of consequences of strong virial / potential energy correlations were also discussed in Paper II. Paper III,\cite{III} published in tandem with this paper, gives further theoretical results on the statistical mechanics and thermodynamics of the hidden scale invariance that characterizes strongly correlating liquids. Paper III also presents results from computer simulations demonstrating that strong virial-potential energy correlations are present even in non-equilibrium situations.

The present paper introduces the new concept of ``isomorphs'' in the state diagram of a strongly correlating liquid and derives a number of isomorph characteristics. The existence of isomorphs distils the properties of strongly correlating liquids into one single concept and its immediate consequences, a concept that is defined without reference to correlations.

In order to recapitulate the definition of a strongly correlating liquid, recall\cite{all87} that for a system of $N$ particles in volume $V$ at temperature $T$, the pressure $p$ is a sum of the ideal gas term $Nk_BT/V$ and a term reflecting the interactions, $W/V$, where $W$ is the so-called virial, i.e.,

\begin{equation}\label{Wdef}
pV \,=\,
Nk_BT+W\,.
\end{equation}
This equation is usually thought of as describing thermodynamic averages, but it also applies for the instantaneous values. The instantaneous ideal-gas pressure term is a function of the particle momenta, giving the $Nk_BT$ term. The instantaneous virial $W$ is the function of particle positions defined\cite{all87} by $W({\bf r}_1,..., {\bf r}_N)\equiv -1/3 \sum_i {\bf r}_i \cdot {\bf \nabla}_{{\bf r}_i} U({\bf r}_1,..., {\bf r}_N)$ where $U({\bf r}_1,..., {\bf r}_N)$ is the potential energy function. If $\Delta U$ is the instantaneous potential energy minus its thermodynamic average and $\Delta W$ the same for the virial, at any given state point the $WU$ correlation coefficient $R$ is defined by (sharp brackets here and henceforth denote equilibrium NVT ensemble averages)

\begin{equation}\label{R}
R \,=\,
\frac{\langle\Delta W\Delta U\rangle}
{\sqrt{\langle(\Delta W)^2\rangle\langle(\Delta U)^2\rangle}}\,.
\end{equation}
By the Cauchy-Schwarz inequality the correlation coefficient obeys $-1\leq R\leq 1$. We define strongly correlating liquids by the condition $R>0.9$.\cite{I} The correlation coefficient is state-point dependent, but for all liquids we have studied by computer simulation\cite{I} $R$ is either above $0.9$ in a large part of the state diagram, or not at all. In all cases the correlation  coefficient quickly decreases when pressure becomes negative, but strongly correlating liquids generally remain so at zero pressure.

Strongly correlating liquids include\cite{ped08a,ped08b,I,II,cos08,cos09} the standard Lennard-Jones (LJ) liquid, the Kob-Andersen binary LJ mixture as well as other binary LJ-type mixtures, a ``dumbbell''  liquid of two different LJ spheres with fixed bond length, a system with exponential repulsion, a seven-site united-atom toluene model, the Lewis-Wahnstr{\"o}m orthoterphenyl model, and an attractive square-well binary mixture. Strongly correlating liquids have simpler physics than liquids in general. This fact has particular significance for the highly viscous phase.\cite{kau48,har76,bra85,gut95,edi96,cha98,ang00,alb01,deb01,bin05,sci05,dyr06} Thus to a good approximation strongly correlating viscous liquids have all eight fundamental frequency-dependent thermoviscoelastic response functions\cite{ell07,bai08c,chr08} given in terms of just one,\cite{ped08b} i.e, they are almost single-parameter liquids in the sense of having dynamic Prigogine-Defay ratio\cite{ell07} close to unity\cite{ped08b,II,bai08c}. Moreover, strongly correlating viscous liquids obey density scaling, i.e., their average relaxation time $\tau$ varies with density $\rho=N/V$ and temperature according to $\tau= F(\rho^\gamma/T)$.\cite{sch08b,tol01,dre03,alb04,cas04,rol05} Even complex systems like biomembranes may exhibit significant correlations for their slow thermodynamic degrees of freedom; this was shown by all-atom computer simulations of five phospholipid membranes which exhibit strong correlations of the energy-volume fluctuations in the NpT ensemble.\cite{ped08c}

When instantaneous values of virial and potential energy are plotted against each other for a strongly correlating liquid in thermal equilibrium at constant volume, an elongated ellipse appears.\cite{ped08a,I,II,cos08} The slope of this ellipse is $\sqrt{{\langle(\Delta W)^2 \rangle }/{\langle(\Delta U)^2 \rangle}}$. As detailed in Sec. IID this quantity, which is weakly state-point dependent, is to a good approximation the exponent $\gamma$ of the density-scaling relation, $\tau= F(\rho^\gamma/T)$.\cite{sch08b,cos09} Thus for a strongly correlating liquid knowledge of the equilibrium fluctuations at one state point provides information about how the relaxation time varies with density and temperature.

What causes strong the $WU$ correlations of some liquids? A hint comes from the well-known fact that an inverse power-law pair potential,\cite{hoo72,hiw74,woo85,bar87,deb99,lan03,spe03,she03,ric05,bra06,cas06,hey07} $v(r)\propto r^{-n}$ where $r$ is the distance between two particles, implies 100\% correlation.\cite{ped08a,II} In this case the slope is $n/3$.\footnote{Potentials consisting of IPL interactions plus an explicit function of volume also have exact isomorphs. Such volume-dependent terms are sometimes used in models of metals, though generally not in conjunction with IPL pair potentials.}  In simulations of the standard LJ liquid we found slopes around 6, corresponding to $n\cong 18$.\cite{ped08a} Although this may seem puzzling given the expression defining the LJ potential $v_{LJ}(r)=4\epsilon[(r/\sigma)^{-12}-(r/\sigma)^{-6}]$, if one wishes to fit the repulsive part of the LJ potential by an inverse power law, an exponent around $18$ is indeed required.\cite{ped08a,II,ben03} The reason is that the attractive $r^{-6}$ term makes the repulsion considerably steeper than the bare repulsive $r^{-12}$ term would imply.

Paper II gave a thorough discussion of the $WU$ correlations with a focus on the standard single-component LJ liquid; this included also a treatment of the (classical) LJ crystal where one finds $0.99<R<1$ at low temperature. According to Paper II the $r$-dependent effective exponent $n$ which controls the correlation is not simply that coming from fitting the repulsive part of the potential, but rather $n^{(2)}(r)\equiv -2-rv'''(r)/v''(r)$. This number is 18-19 around the LJ minimum. In fact, the LJ potential may here be fitted very well with an ``extended inverse power-law'' (eIPL) potential,\cite{II,III} a potential of the form $v_{\rm LJ}(r)\cong A r^{-n}+B+Cr$ with $n$ of order $18$. For this potential $n^{(2)}(r)=n$. At constant volume the linear term contributes little to the virial and potential-energy fluctuations: When one nearest-neighbor interatomic distance increases, another decreases in such a way that their sum remains almost constant.\cite{II} This means that virtually correct canonical probabilities are arrived at by using the inverse power-law (IPL) approximation, an observation which inspired us to the below isomorph definition. 

For an IPL liquid several quantities are invariant along the curves in the phase diagram given by $\rho^{n/3}/T={\rm Const.}\,$ Paper III\cite{III} summarizes the thermodynamic IPL invariants, which include Helmholtz free energy over temperature, excess entropy, average potential energy over temperature, isothermal bulk modulus over density times temperature, and virial over temperature. In reduced units the dynamics of an IPL liquid is also invariant along the $\rho^{n/3}/T={\rm Const.}$ curves.\cite{III} The present paper shows that some IPL invariants give rise to general ``isomorph invariants'' of strongly correlating liquids. Not all IPL invariants generalize, however, and e.g. the equation of state of a strongly correlating liquid is usually poorly represented by the IPL approximation.\cite{II} 

We demonstrate below several implications of one single assumption: the existence of curves in the phase diagram on which for any two state points there is a one-to-one correspondence between their respective microscopic configurations, such that corresponding configurations have identical configurational NVT canonical probabilities. These curves in the state diagrams are referred to as \emph{isomorphs}. Section II defines isomorphs and summarizes their properties classified into thermodynamic, structural, equilibrium dynamic, and aging properties. Most isomorph properties come in the form of isomorph invariants. In Sec. II we also discuss how to identify isomorphic curves in the state diagram. Section III presents results from computer simulations of (mainly) the Kob-Andersen binary LJ mixture, validating some of the isomorph predictions. Section IV relates the isomorph concept to selected topics of current liquid state theory and experiment. Section V gives a brief conclusion.

\section{Isomorphs}

This section introduces the concept of isomorphs in the state diagram of a strongly correlating liquid. Although the definition of an isomorph refers neither to IPL potentials nor to strong $WU$ correlations, only strongly correlating liquids have isomorphs. This is because the existence of isomorphs reflects the hidden scale invariance that characterizes strongly correlating liquids.\cite{II,III,sch08b}

\subsection{Isomorph definition}

Assuming that the origin of the coordinate system is centred in the liquid, for any microscopic configuration $({\bf r}_1,\, ...\, , {\bf r}_N)$ of a thermodynamic state point with density $\rho$, the ``reduced'' (dimensionless) coordinates are defined by

\begin{equation}\label{tilderdef}
\tilde{\bf r}_i\,\equiv\,\rho^{1/3} {\bf r}_i \,.
\end{equation}
Using reduced coordinates corresponds to switching to a scaled coordinate system where the density is unity. We term a microscopic configuration {\it physically relevant} with respect to a given thermodynamic state point if the configuration's contribution to the partition function at that state point is not a priori negligible. For instance, no configurations where all particles occupy the left half of the system's volume are physically relevant for ordinary liquid states. 

State points (1) and (2) with temperatures $T_1$ and $T_2$ and densities $\rho_1$ and $\rho_2$ are {\it isomorphic} if they obey the following: Whenever two of their physically relevant microscopic configurations $({\bf r}_1^{(1)},\, ...\, , {\bf r}_N^{(1)})$ and $({\bf r}_1^{(2)},\, ...\, , {\bf r}_N^{(2)})$ have identical reduced coordinates (i.e., $\tilde{\bf r}_i^{(1)}=\tilde{\bf r}_i^{(2)}$), they have proportional configurational NVT Boltzmann factors:

\begin{equation}\label{isodef}
e^{-U({\bf r}_1^{(1)},\, ...\, , {\bf r}_N^{(1)})/k_BT_1}\, =\, C_{12}\,e^{-U({\bf r}_1^{(2)},\, ...\, , {\bf r}_N^{(2)})/k_BT_2}\,.
\end{equation}
It is understood that the constant $C_{12}$ depends only on the state points (1) and (2), not on the microscopic configurations. {\it Isomorphic curves} in the state diagram are defined as curves on which any two state point are isomorphic. 

An IPL liquid with interactions scaling with distance $\propto r^{-n}$ trivially obeys Eq. (\ref{isodef}) with $C_{12}=1$ for states with $\rho_1^{n/3}/T_1=\rho_2^{n/3}/T_2$. No other systems obey Eq. (\ref{isodef}) rigorously, but we show in Sec. III from simulations of a strongly correlating liquid that the existence of isomorphs is a good approximation. Although only IPL liquids have exact isomorphs, we shall say that a liquid ``has isomorphs'' if isomorphic curves exist to a good approximation in the liquid's phase diagram. 

The isomorph definition does not refer to $WU$ correlations, but only strongly correlating liquids have isomorphs. The Appendix proves in detail that all strongly correlating liquids have isomorphs and vice versa. It is illuminating here to briefly sketch why the existence of isomorphs for a liquid implies that it must be strongly correlating: Consider a liquid with two isomorphic state points that are infinitesimally close to each other. If $\delta$ represents the variation between two infinitesimally close microscopic configurations of the state points with same reduced coordinates, taking the logarithm of Eq. (\ref{isodef}) leads to $\delta (U/T)={\rm Const.\,}$ where the constant is infinitesimal. The differentiation leads to a relation of the form $\delta U = (da)U+db$ where $da$ and $db$ are infinitesimal. Since $\delta U$ is the potential energy difference between two microscopic configurations differing by $\delta \br_i\propto\br_i$, one has $\delta U\propto  W$ (Appendix A). Altogether we get $W=AU+B$ for some constants $A$ and $B$. This implies 100\% correlation of the $WU$ fluctuations -- recall, however, that the existence of isomorphs is itself an approximation. Thus liquids having isomorphs to a good approximation must be strongly correlating. 

A number of properties characterize isomorphic curves in the state diagram of a strongly correlating liquid. Most isomorph properties come in the form of isomorph invariants. These are consequences of the fact that for any two isomorphic state points there is a one-to-one correspondence between the state points' physically relevant microscopic configurations, such that corresponding configurations have the same NVT canonical probabilities. This one-to-one correspondence motivates the name isomorph (``same form''), which is fundamental throughout mathematics. Here two objects are termed isomorphic if they are structurally equivalent, i.e., if a structure-preserving bijective mapping between them exists. In physics and chemistry isomorphic crystals by definition have symmetry groups that are mathematically isomorphic -- such crystals have the same structure but different constituents.

\subsection{Isomorph properties}

Before detailing the consequences of the isomorph definition Eq. (\ref{isodef}) we briefly recall well-known facts of the statistical mechanics of classical liquids in the NVT canonical ensemble.\cite{lan80,all87,cha87,rei98,han05} The Helmholtz free energy $F$ is the sum of an ideal gas term and an ``excess'' free energy term reflecting the molecular interactions, $F=F_ {\rm id}+F_ {\rm ex}$. The first term is the free energy of an ideal gas at the same density and temperature, $F_ {\rm id}=-Nk_BT\ln(\rho\Lambda^3)$ where $\Lambda=h/\sqrt{2\pi mk_BT}$ is the thermal de Broglie wavelength. The excess free energy is given by

\begin{equation}\label{fex}
e^{-F_ {\rm ex}/k_BT}\,=\,
\int \frac{d{\bf r}_1}{V}...\frac{d{\bf r}_N}{V}e ^{-U(\bR)/k_BT}\,.
\end{equation}
The integral involves only the configurational degrees of freedom. This separation of the configurational degrees of freedom from the momenta is assumed throughout this paper. Thus when we refer to the canonical ensemble and canonical probabilities, only the configurational part of phase space is implied. Likewise, the microcanonical ensemble below refers to the uniform probability distribution on the constant potential energy surface in configuration space.

The configuration space probability distribution normalized to the above dimensionless integral is given by

\begin{equation}\label{pdef}
P(\bR)\,=\,
e^{- [U(\bR)-F_ {\rm ex}]/k_BT}\,.
\end{equation}
The excess entropy $S_ {\rm ex}$ is defined by $S_ {\rm ex} =-\partial F_ {\rm ex}/\partial T$. Since the entropy in the canonical ensemble is generally given by $-k_B\langle\ln P\rangle$, we have

\begin{equation}\label{sex}
S_ {\rm ex}\,=\,
-k_B\int \frac{d{\bf r}_1}{V}...\frac{d{\bf r}_N}{V}P(\bR)\ln P(\bR)\,.
\end{equation}
The inequality $-P\ln P \leq 1-P$ implies that $S_ {\rm ex}$ is always negative, a fact that is physically obvious since any liquid is more ordered than an ideal gas at the same volume and temperature.

Most isomorph invariants are consequences of two fundamental isomorph properties, Eqs. (\ref{pinv}) and (\ref{uinv}) below. Equation (\ref{isodef}) implies that the normalized reduced-coordinate probability distribution is invariant along an isomorph:

\begin{equation}\label{pinv}
\tilde P(\tR)\,\, {\rm is\,an\,isomorph\,invariant}\,.
\end{equation}
The notation $\tilde P$ is introduced to distinguish from the $P$ of Eq. (\ref{pdef}); $\tilde P$ is normalized via $\int \tilde P (\tilde{\bf r}_1, ... ,\tilde{\bf r}_N) d\tilde{\bf r}_1 ...d\tilde{\bf r}_N =1$ which implies $ \tilde P(\tR) =N^{-N}P(\bR)$. The second fundamental isomorph property is that an isomorph $I$ is characterized by two functions, $f_I(\tR)$ and $g(Q)$ where $Q$ denotes the state point, such that for any physically relevant microscopic configuration of state point $Q$

\begin{equation}\label{uinv}
U(\bR;\rho)\,=\,
k_BT\,f_I(\tR)\,+\,g(Q)\,.
\end{equation}
This follows from the isomorph definition Eq. (\ref{isodef}). In this formulation the potential energy function is formally regarded as density dependent, a dependence that is to reflect the fact that Eq. (\ref{uinv}) only applies for physically relevant microscopic configurations, i.e., configurations that fill out the volume.

Before deriving the isomorph properties we note two multidimensional geometric isomorph characterizations. Recall that the potential energy landscape is the graph of the potential energy function, i.e., a subset of $R^{3N+1}$. The first geometric property is that isomorphic state points have potential energy landscapes which, when restricted to the physically relevant states, are identical except for a vertical displacement and scaling by the inverse temperature, and a horizontal scaling to unit density. This follows from the isomorph definition Eq. (\ref{isodef}), and this is what Eq. (\ref{uinv}) expresses. The second geometric isomorph characterization relates to the hypersurface $\Omega$ in $R^{3N}$ where the potential energy equals the average potential energy of the state point in question: In reduced coordinates this ``constant potential energy hypersurface'', denoted by  $\tilde\Omega$, is invariant along an isomorph. This follows from the first geometric property in conjunction with the isomorph definition Eq. (\ref{isodef}). For details, please consult Appendix A, which gives detailed proofs that a liquid is strongly correlating if and only if it has isomorphs, and that this happens if and only if the liquid has curves in the state diagram along which $\tilde\Omega$ is invariant.

\subsubsection{Thermodynamics}

A number of thermodynamic quantities are invariant along isomorphic curves in the state diagram of a strongly correlating liquid.

\begin{itemize}

\item 1a. {\it The excess entropy $S_ {\rm ex}$ is invariant along an isomorph.} 
Equation (\ref{sex}) implies that $S_ {\rm ex}=-k_B\int \tilde P \ln\tilde P d\tilde{\bf r}_1 ...d\tilde{\bf r}_N+{\rm Const.}$, from which property 1a follows because of the invariance of $\tilde P$. Property 1a may also be derived by referring to the microcanonical ensemble where the excess entropy is $k_B$ times the logarithm of the area of $\tilde\Omega$: Because $\tilde\Omega$ is an isomorph invariant, so is the excess entropy.

\item 1b. {\it The configurational entropy $S_ {\rm conf}$ is invariant along an isomorph.} 
The term ``configurational entropy'' -- not to be confused with $S_ {\rm ex}$ -- is used here in the sense of Adam and Gibbs and subsequent workers, who related $S_ {\rm conf}$ to the temperature dependence of viscous liquids' average relaxation time.\cite{ada65,edi96,ang00} The configurational entropy is $k_B$ times the logarithm of the ``density of states'' of potential energy minima, the so-called inherent states, evaluated at the state point's average inherent state energy.\cite{deb01,sci05} Property 1b follows from the identity of the scaled potential energy landscapes of two isomorphic state points.

\item 1c. {\it When a liquid is heated slowly along an isomorph, the measured specific heat equals the ideal-gas specific heat.}
Since the specific heat is $dS/d\ln T$ and property 1a implies $dS_ {\rm ex}=0$ along an isomorph, the measured specific heat equals the specific heat that would be measured for an ideal gas subjected to the same process.

\item 1d. {\it The isochoric specific heat is invariant along an isomorph.} 
If the liquid's (extensive) excess isochoric specific heat is denoted by $\ccv$, in terms of the variable $X=U/k_BT$ Einstein's expression $\ccv=\langle(\Delta U)^2\rangle/k_BT^2$ becomes $\ccv =k_B \langle(\Delta X)^2\rangle$. Equation (\ref{uinv}) implies that for two isomorphic state points their  microscopic configurations with same reduced positions obey $X_1 =X_2+{\rm Const.}$, i.e., for the fluctuations $\Delta X_1 =\Delta X_2$. Since $\Delta X$ depends only on the reduced coordinates and since $\tilde P (\tR)$ is isomorph invariant, it follows that the excess isochoric specific heat is an isomorph invariant. This implies invariance of the full $\ccv$, because the contribution to $\ccv$ from the momentum degrees of freedom is state-point independent.

\end{itemize}

\subsubsection{Structure}

Particle distribution functions are generally invariant along an isomorph when quoted in reduced coordinates.

\begin{itemize}

\item 2a. {\it Scaled radial distribution function(s) -- as well as higher-order equilibrium particle probability distributions -- are invariant along an isomorph}. Property 2a follows from Eq. (\ref{pinv}); note that it applies for liquids with any number of different types of particles.

\item 2b. {\it The multiparticle entropies $S_2, S_3, ...$ are invariant along an isomorph.}
The equilibrium particle distributions give rise to n-particle entropies\cite{mulen,bar89} contributing to the total entropy as follows $S_ {\rm ex}=S_2+S_3+...$. In terms of the radial distribution function for a system of identical particles $g(r)$, the pair-correlation contribution $S_2$ is given by $S_2/N=-(\rho k_B/2)\int d\br\left[g(r)\ln g(r) +1 - g(r)\right]$. When this expression, as well as the more involved expressions defining $S_3$, etc., are rewritten in terms of reduced coordinates, it becomes clear that property 2b is a consequence of Eq. (\ref{pinv}). We considered here only the case of identical particles, but property 2b holds for systems with any number of different particles.

\end{itemize}

\begin{table}[ht]
\begin{center}
 \begin{tabular}{l cc}
\hline \hline
\textbf{Quantity}& \textbf{Newtonian dynamics} & \textbf{Brownian dynamics} \\ 
\hline
\textrm{Energy unit ($E_0$)} & $k_BT$ & $k_BT$\\
\textrm{Length unit ($l_0$)} & $\rho^{-1/3}$ & $\rho^{-1/3}$\\
\textrm{Time unit ($t_0$)} & $\frac{\sqrt{m/k_BT}}{\rho^{1/3}}$ &$\frac{1}{\rho^{2/3}\mu k_BT}$\\
\hline
\textrm{Diffusion constant:} & &\\
$\tilde D=D/(l_0^2/t_0)$ & $\tilde D=(\rho^{1/3}\sqrt{m/k_BT})\,D$ & $\tilde D=\frac{1}{\mu k_BT}\,D$\\
\hline
\textrm{Viscosity:} & &\\
$\tilde \eta=\eta/(E_0t_0/l_0^3)$ & $\tilde\eta=\frac{1}{\rho^{2/3}\sqrt{mk_BT}}\,\eta$ 
& $\tilde\eta=\frac{\mu}{\rho^{1/3}}\,\eta$\\
\hline
\textrm{Heat conductivity:} & &\\
$\tilde\kappa=\kappa/(k_B/l_0t_0)$ & $\tilde\kappa=\frac{\sqrt{m/k_BT}}{\rho^{2/3}k_B}\,\kappa$ & $\tilde\kappa=\frac{1}{\rho \mu k_B^2T}\,\kappa $\\
\hline
\end{tabular}
\end{center}
\caption{Reduced units. The energy and length units refer to state-point properties only, the time unit depends also on the dynamics. Once these three fundamental units have been defined, transport and other properties have uniquely defined dimensionless versions, denoted by a tilde. The table gives three examples of such properties.}\label{red_table}
\end{table}

\subsubsection{Equilibrium dynamics}

As for the static isomorph invariants, dynamic invariants also derive from the fact that for all physically relevant microscopic configurations the potential energy landscapes of isomorphic state points are identical -- except for additive constants and overall scalings that do not affect the reduced dynamics. For reference below, Table \ref{red_table} summarizes the definition of the basic units and gives examples of some of the corresponding reduced quantities.

\begin{itemize}

\item 3a. {\it Both NVE and NVT Newtonian dynamics are isomorph invariant when described in reduced units.}
Consider first standard energy-conserving Newtonian dynamics, the NVE ensemble. If the mass of particle $i$ is $m_i$, Newton's second law is $m_i\ddot\br_i=\bF_i$ where $\bF_i=-\nabla_{{\bf r}_i} U$ is the force on the i'th particle. We rewrite this in terms of reduced units as follows. If the average particle mass is $m$, the reduced mass of the i'th particle is defined by $\tm_i=m_i/m$. The reduced potential energy $\tU$ is defined by $\tU=U/k_BT$ and the reduced force $\tF_i$ by $\tF_i=-\nabla_{\tilde{\bf r}_i} \tU$. We also define a reduced time, $\tti=t/t_0$, where $t_0=\rho^{-1/3}\sqrt{m/k_BT}$. In terms of these reduced variables Newton's second law becomes $\tm_i{\ddot\tr}_i=\tF_i$. The isomorph invariance now follows from Eq. (\ref{uinv}), which implies that the reduced force is the same function of the reduced particle positions for all state points on a given isomorph. Proceeding to Newtonian dynamics in the NVT ensemble realized via the Nosé-Hoover thermostat,\cite{nos84,hoo85} recall that the forces here have the additional term $-\zeta m_i\dot\br_i$ with a ``friction constant'' $\zeta$ obeying $\dot\zeta=(K/K_0-1)/\tau_0^2$, where $K$ is the kinetic energy, $K_0$ its average, and $\tau_0$ the thermostat time constant. These equations become isomorph invariant when rewritten in terms of the same reduced units as the NVE Newtonian equations, if the thermostat time constant is adjusted to be the same in reduced units; otherwise, only the long-time NVT dynamics is isomorph invariant.

\item 3b. {\it  Brownian dynamics is isomorph invariant when described in reduced units.}
The Brownian (Langevin) equations of motion are first-order stochastic equations. These equations obey detailed balance, ensuring consistency with the canonical ensemble. The Brownian equation of motion is $\dot\br_i=\mu\bF_i+\bm{\xi}(t)$ where $\mu$ is the ``mobility'' (velocity/force) and $\bm{\xi}(t)$ is a Gaussian white-noise term characterized by $\langle\bm{\xi}_m(t)\bm{\xi}_n(t')\rangle=2\mu k_BT\delta_{mn}\delta(t-t')\,\,(m,n=1,2,3)$. The path-probability functional is given\cite{ris89} by 
$P\propto\exp[-1/(4\mu k_BT)\sum_i\int_{-\infty}^{\infty}(\dot\br_i-\mu\bF_i)^2dt]$. We rewrite this in terms of reduced variables with $\tU=U/k_BT$ as above, but a reduced time that is now defined via $\tti=t/t_0$ where $t_0=\rho^{-2/3}/\mu k_BT$. This leads to $P\propto\exp[-1/4\sum_i\int_{-\infty}^{\infty}({\dot{\tr}}_i-\tilde\bF_i)^2d\tti]$. Since the reduced force is isomorph invariant for microscopic configurations with same reduced coordinates, it follows that reduced-time Brownian dynamics is isomorph invariant.

\item 3c. {\it  Normalized time-autocorrelation functions -- as well as normalized higher-order time correlation functions -- are invariant along an isomorph when quoted in reduced units.}
Consider the time-autocorrelation function or higher-order time-correlation functions of some variable $A$ referring to constant volume dynamics. Properties 3a and 3b imply that for both Newtonian and Brownian dynamics time-autocorrelation functions of $A$ are invariant as functions of the reduced time if they are normalized by dividing by $\langle A^2\rangle$. Normalization of any higher-order time-correlation function similarly  makes it isomorph invariant as function of the reduced times.

\item 3d. {\it Average relaxation times are isomorph invariant when quoted in reduced units.}
For any variable $A$ with zero mean a generic definition of its average relaxation time is $\tau_A=\int_0^\infty\langle A(0)A(t)\rangle dt/\langle A^2\rangle$. In reduced units this becomes
$\tilde\tau_A=\int_0^\infty\langle \tilde A(0)\tilde A(\tti)\rangle d\tti/\langle \tilde A^2\rangle$. By property 3c this expression is isomorph invariant.

\item 3e. {\it Reduced transport coefficients like the diffusion constant, the viscosity, etc., are invariant along an isomorph.}
By the fluctuation-dissipation (FD) theorem the diffusion constant is given by $D=\int_0^\infty\langle v_x(0)v_x(t)\rangle dt$ where $v_x$ is the $x$ component of a particle's velocity. The reduced diffusion constant $\tilde D$ is defined by $\tilde D=(\rho^{1/3}\sqrt{m/k_BT})D$ for Newtonian and $\tilde D=D/(\mu k_BT)$ for Brownian dynamics (Table \ref{red_table}). In both cases one has $\tilde D=\int_0^\infty\langle \tilde v_x(0)\tilde v_x(t)\rangle d\tti$, implying that $\tilde D$ is an isomorph invariant because both dynamics are isomorph invariant. Similarly, if $\eta$ is the viscosity, the reduced viscosity is defined by $\tilde\eta=(\rho^{-2/3}/\sqrt{mk_BT})\eta$ for Newtonian dynamics and $\tilde\eta=(\mu\rho^{-1/3})\eta$ for Brownian dynamics. When rewritten as a reduced-time integral over the reduced shear-stress autocorrelation function, the required result follows. Similar results apply for the heat conductivity and other DC transport coefficients.

\item 3f. {\it $G_\infty/T \rho$ is invariant along an isomorph where $G_\infty$ is the instantaneous shear modulus.}
If $S_{xy}=\sum_i x_i F_{i,y}$ where $x_i$ is the x-coordinate of the i'th particle and $F_{i,y}$ is the y-component of the force acting on it, the FD expression for the instantaneous shear modulus\cite{zwa65} is $G_\infty=\rho k_BT+\langle S_{xy}^2\rangle/V k_BT$. In terms of reduced variables one has $x_i F_{i,y}=-k_BT\tilde x_i \partial\tU/\partial\tilde y_i$, leading to 
$G_\infty/\rho k_BT=1+\langle (\sum_i \tilde x_i \partial\tU/\partial\tilde y_i )^2\rangle/N$. The required isomorph invariance now follows from Eq. (\ref{pinv}).

\end{itemize}

\subsubsection{Aging}

Not all isomorph properties come in the form of invariants. This section discusses a different type of consequence of the existence of isomorphs for strongly correlating liquids. It is assumed that the externally controlled variables are volume and temperature.

\begin{itemize}

\item 4a. {\it A jump between two isomorphic state points starting from equilibrium takes the system instantaneously to equilibrium.}
This is because the normalized Boltzmann probability factors are identical for the two systems. Thus isomorphs are predicted to be a kind of ``wormholes'' in the state diagram along which one can jump instantaneously from equilibrium to equilibrium, even when the states are characterized by long relaxation times.

\item 4b. {\it Jumps from any two isomorphic equilibrium state points to a third state point lead to the same aging behavior for all physical quantities.}
Note first that a jump between two arbitrary state points starting from equilibrium, $1\rightarrow 3$, has the same relaxation pattern as the $2\rightarrow 3$ jump, where state point (2) is isomorphic with state point (1) and has the same density as state point (3): Suppose that instead of the $1\rightarrow 3$ jump we first impose the isomorphic jump to state point (2) and then, immediately thereafter, jump to state point (3). On the one hand, the system will never ``register'' it spent a tiny amount of time at state point (2). On the other hand, the $1\rightarrow 2$ jump took the system instantaneously to equilibrium at state point (2) (property 4a). Consequently, the $1\rightarrow 3$ and $2\rightarrow 3$ jumps must have the same relaxation towards equilibrium for all physical quantities. It now follows that if state point (1) is replaced by an isomorphic state point (1'), the $1\rightarrow 3$ and $1'\rightarrow 3$ relaxations towards equilibrium are identical.

\end{itemize}

A concise way of summarizing the aging properties of strongly correlating liquids is that {\it isomorphic state points are equivalent during any aging scheme.}

\subsection{IPL invariants and general isomorph invariants}

As mentioned, an IPL liquid has exact isomorphs. These are the curves in the state diagram given by the equation $\rho^{n/3}/T={\rm Const.}\,$  All above isomorph properties apply exactly to IPL liquids (single- or multicomponent, as long as the exponent is the same for all particles). Only some of the IPL invariants along the curves given by $\rho^{n/3}/T={\rm Const.}$, however, give rise to general isomorph invariants. Examples of IPL invariants that do not generalize are: $F_ {\rm ex}/T$, $U/T$, $W/T$, the excess pressure coefficient $\bv=(\partial (W/V)/\partial T)_V$, and $K_T/T\rho$ where $K_T$ is the isothermal bulk modulus. One way of determining which IPL thermodynamic properties give rise to general isomorph invariants is to use the approximate equation for the excess free energy derived in Paper III.\cite{III} Here it was shown that the hidden scale invariance of a strongly correlating liquid implies that one may write $F_ {\rm ex}(V,T)=f(V)+Nk_BT\phi(\rho^\gamma/T)$ to a good approximation. For an IPL liquid $f(V)=0$ which, however, does not apply generally. The existence of the $f(V)$ term implies that IPL properties involving volume derivatives do not give rise to general isomorph invariants.

\subsection{Identifying isomorphs}

How to identify the isomorphic curves in a strongly correlating liquid's state diagram? As we just saw, for an IPL liquid the answer is simple: All static and dynamic IPL invariants refer to the curves given by $\rho^\gamma/T={\rm Const.}$ where $\gamma=n/3$. If the virial and potential energy equilibrium fluctuations from their average values obey $\Delta W=\gamma \Delta U$, the number $\gamma$ may be expressed in terms of equilibrium fluctuation averages in three simple ways: 
$\gamma=\langle\Delta W\Delta U\rangle/\langle(\Delta U)^2\rangle=\sqrt{\langle(\Delta W)^2\rangle/\langle(\Delta U)^2\rangle}
=\langle(\Delta W)^2\rangle/\langle\Delta W\Delta U\rangle$. This applies only for a 100\% correlating liquid, however, i.e. for an IPL liquid. For a general strongly correlating liquid there are three different corresponding gammas,

\begin{equation}\label{gammadef}
\gamma_1\,=\,\frac{\langle\Delta W\Delta U\rangle}{\langle(\Delta U)^2\rangle}\,,\,\,\,\,
\gamma_2\,=\,\sqrt{\frac{\langle(\Delta W)^2\rangle}{\langle(\Delta U)^2\rangle}}\,,\,\,\,\,
\gamma_3\,=\,\frac{\langle(\Delta W)^2\rangle}{\langle\Delta W\Delta U\rangle}\,.
\end{equation}
It follows from the definition of the correlation coefficient $R$ (Eq. (\ref{R})) that 

\begin{equation}\label{gammaid}
\gamma_1\,=\,R\gamma_2\,=\,R^2\gamma_3\,.
\end{equation}
Thus $\gamma_2$ is the geometric mean of $\gamma_1$ and $\gamma_3$, and

\begin{equation}\label{gammarel}
\gamma_1\leq \gamma_2\leq \gamma_3\,.
\end{equation}
Although for any strongly correlating liquid the three gammas are quite similar, the question is which gamma to use to identify the isomorphs? The answer is that there is no unique gamma. To see this, note that in complete generality any quantity $q$ of any liquid defines a state-point dependent exponent $\gamma_q(Q)$ with the following property: An infinitesimal change away from state point $Q$ conserves $q$ whenever the quantity $\rho^{\gamma_q(Q)}/T$ is kept constant (thus  $\gamma_q(Q)=(\partial\ln T/\partial\ln\rho)_q (Q)$). Strongly correlating liquids are characterized by the particular property that these $\gamma$'s are all very similar for the isomorph invariants. 

In the simulations reported in the next section we used the excess entropy's ``density scaling exponent'' $\gamma$ derived as follows. Along a configurational adiabatic curve $0=dS_{\rm ex}=(\partial S_{\rm ex}/\partial V)_TdV+(\partial S_{\rm ex}/\partial T)_VdT$. The volume-temperature Maxwell relation for the configurational degrees of freedom implies $(\partial S_{\rm ex}/\partial V)_T=(\partial (W/V)/\partial T)_V $. Thus a configurational adiabat is characterized by $(d\ln\rho) (\partial W/\partial T)_V =(d\ln T) T (\partial S_{\rm ex}/\partial T)_V=(d\ln T) (\partial U/\partial T)_V$, i.e., 

\begin{equation}\label{gammaS}
\gamma
\,=\,\left(\frac{d\ln T}{d\ln \rho}\right)_{S_{\rm ex}}
\,=\, \frac{\left(\frac{\partial W}{\partial T}\right)_V}{\left(\frac{\partial U}{\partial T}\right)_V}
\,=\, \frac{\langle\Delta W\Delta U\rangle}{\langle(\Delta U)^2\rangle}\,.
\end{equation}
This is $\gamma_1$ of Eq. (\ref{gammaS}); the last equality is a standard thermodynamic fluctuation identity derived, e.g., in Appendix B of Paper I. Since it is convenient to think of the density scaling exponent as a generic quantity, we will not refer to it as $\gamma_1$, but simply as $\gamma$. Note that $\gamma=(\partial W/\partial U)_V$, which implies that $\gamma$ is the slope of the lines of average virial versus average potential energy at constant density in the $UW$ plot (compare Fig. 4 in Paper I). 

We propose to use the excess entropy's density scaling exponent because, on the one hand, it rigorously reproduces property 1a and, on the other hand, it may be calculated from equilibrium fluctuations at any given state point. It may be shown that for a given infinitesimal density change the optimal temperature change leading to a mean-square best fit of the logarithm of Eq. (\ref{isodef}) is obtained by using this $\gamma$ (Appendix B). Finally, note that this $\gamma$ is also the one suggested by taking the ratio of Eqs. (28) and (27) in Paper III: $\gamma=\bv/\cv$ where $\bv=(\partial (W/V)/\partial T)_V$ is the excess pressure coefficient and $\cv$ is the excess isochoric specific heat per unit volume.

It is now clear how to step-by-step map out an isomorph which is realized as a configurational adiabat: At any given state point one calculates $\gamma$ from the equilibrium fluctuations using of Eq. (\ref{gammaS}). For a slight density change one then determines the temperature that keeps $\rho^\gamma/T$ unchanged from the initial state point. This identifies a new state point that is isomorphic with the initial one. At the new state point a new value of $\gamma$ is calculated, etc. We used this method in the simulations reported below, but found almost as good agreement with predictions using $\gamma_2$ ($\gamma_3$ was not tested). 

In the below simulations, as well as in previous simulations, we generally found only weak state-point dependence of the density scaling exponent $\gamma$. As shown in Appendix B, for a strongly correlating liquid any variation of $\gamma$ with state point can only come via a density dependence: $\gamma=\gamma(\rho)$. This result is derived in the ``isomorph approximation'' where there is identity of the curves in the phase diagram of constant excess entropy and those of constant isochoric specific heat. Our simulations are consistent with this result in the sense that $\gamma$ varies significantly less along isochores than otherwise throughout the state diagram. The conclusion is that, in the isomorph approximation, any isomorph invariant is a function of $\rho^{\gamma(\rho)}/T$, i.e., it may be written as a function of a variable of the form $e(\rho)/T$. This is the original recipe for collapsing measurements at different pressures and temperatures, which was proposed by Alba-Simionesco and her coworkers in their pioneering paper on density scaling.\cite{alb04}

\section{Computer simulations}

This section presents results from computer simulations investigating some of the predicted isomorph properties. The purpose is to document the existence of isomorphs for a typical strongly correlating liquid. Results are reported for molecular dynamics simulations of the standard Kob-Andersen\cite{LJ,KA} 80:20 binary Lennard-Jones liquid (KABLJ) with $N=8000$ particles simulated using the Gromacs package.\cite{ber95,lin01} The KABLJ system is a  strongly correlating liquid\cite{ped08a,I} which is easily  supercooled without crystallizing.\cite{tox09} Its density scaling exponent $\gamma$ varies slightly with state point, but at low and moderate pressure and temperature $\gamma$ stays between 5 and 6 (at extremely high pressure and temperature it converges to the value 4, which a purely repulsive $r^{-12}$ potential would imply). The simulations were performed in the NVT ensemble using the Nos\'{e}-Hoover thermostat with characteristic time 0.5 in MD units. Collections of isomorphic state points were identified as described above.

\subsection{Direct isomorph check}

\begin{figure}
\begin{center}
\includegraphics[width=0.5\textwidth]{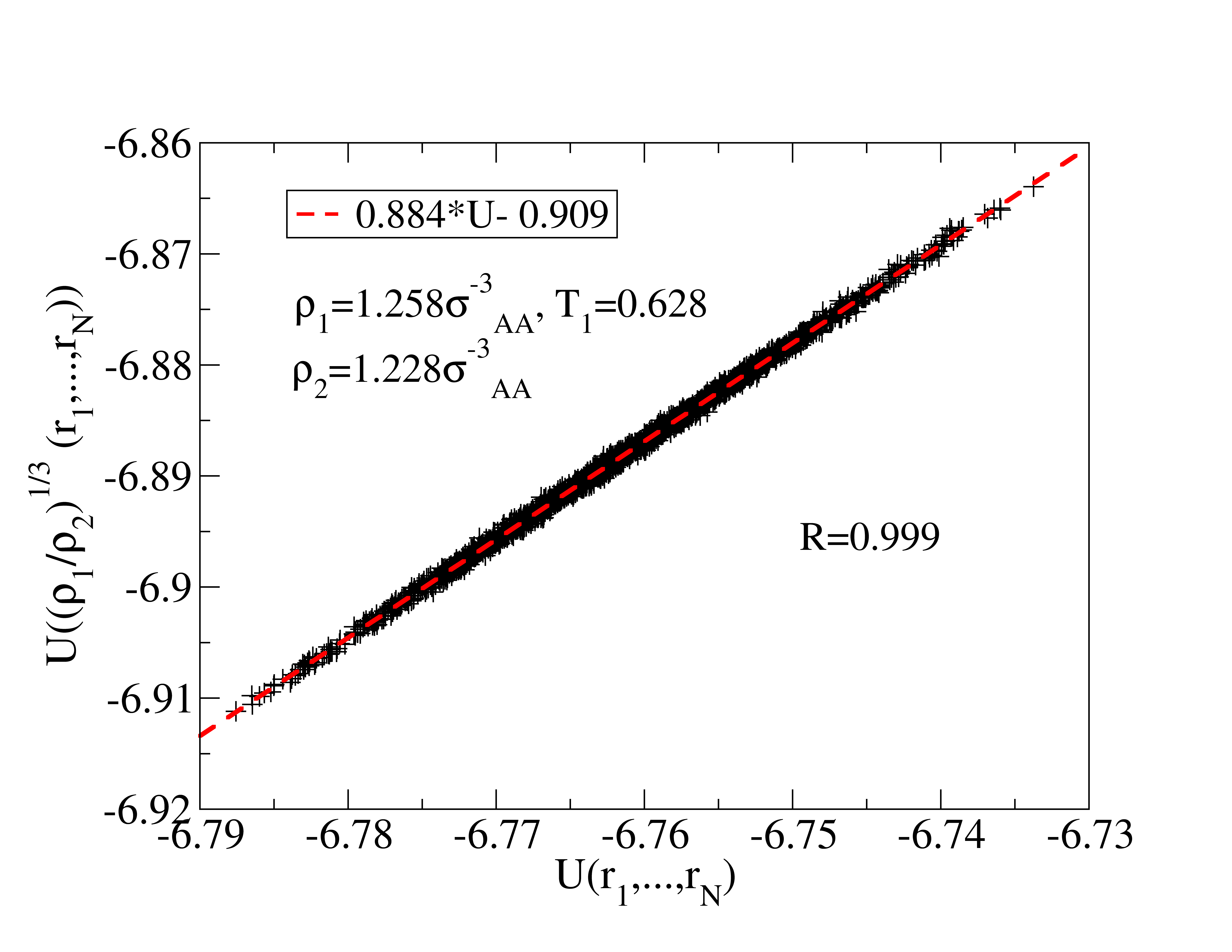}
\end{center}
\caption{\footnotesize{Direct check of the isomorph condition Eq. (\ref{isodef}) for the Kob-Andersen binary Lennard-Jones (KABLJ) liquid with 8000 particles. The molecular dynamics algorithm generates a time sequence of equilibrium microscopic configurations at the state point (1) given by $\rho_1=1.258$ and $T_1=0.628$ in standard LJ units. For each configuration the potential energy is plotted on the x-axis, while the y-axis gives the potential energy of the same microscopic configuration scaled to density $\rho_2=1.228$. Clearly these two quantities are highly correlated, which is the requirement for an isomorph. The slope of $0.884$ implies that state point $(\rho_2,T_2)$ is isomorphic with state point $(\rho_1,T_1)$ if $T_2=0.884T_1=0.555$. In comparison, the exponent of Eq. (\ref{gammaS}) evaluated from the fluctuations at state point (1) is $\gamma=5.018$, and the predicted isomorph temperature at state point (2) calculated from the requirement of keeping $\rho^\gamma/T$ constant is $T_2=0.556$. The non-zero offset of the best fit line (of $-0.909$) reflects the fact that the constant $C_{12}$ in Eq. (\ref{isodef}) differs from unity.}} \label{fig:figure0}
\end{figure} 

It is possible to check directly the proportionality between Boltzmann factors which defines an isomorph. This is done as follows. At one state point of the equilibrium KABLJ liquid the simulation generates a time sequence of microscopic configurations. We now ask: For a given density change, does a temperature exist at which the new density state point is isomorphic to the initial one? If yes, what is this temperature? These questions are answered by plotting the potential energy of each microscopic configuration against the potential energy of the same configuration scaled to the new density. In Fig. \ref{fig:figure0} density was decreased by 2.4\%, which corresponds to a decrease of the relaxation time by more than a factor of four if temperature is not changed. The two potential energies are 99.9\% correlated. This shows that a temperature does exist at which the isomorph condition Eq. (\ref{isodef}) is fulfilled to a very good approximation. When  the density change goes to zero, the correlation goes to 100\% of course, but the high correlation is nevertheless noteworthy given the fact that the slope of the stretched oval in the figure is not very close to unity (the slope is $0.884$). The interpretation of the slope is that the two state points are isomorphic if the temperature at the new density is $0.884$ times the old temperature. -- Appendix C details how the correlation coefficient of Fig. \ref{fig:figure0} relates to the standard $WU$ correlation coefficient of Eq. (\ref{R}).

One might think that Fig. \ref{fig:figure0} proves that all the isomorph invariants apply to a good approximation for the KABLJ liquid, but this is not necessarily the case. Consider for instance the dynamics. This becomes increasingly barrier dominated as temperature is lowered,\cite{heu08} and the increasingly unlikely event of a barrier transition determines the relaxation time. Even if almost all microscopic configurations of significant weight in the canonical ensemble obey the isomorph definition Eq. (\ref{isodef}) very well, this does not ensure that the barriers scale in the same way. Consequently we need more simulations to validate the isomorph concept.

\subsection{Equilibrium properties: Statics and dynamics}

\begin{figure} 
\begin{center}
\includegraphics[width=0.3\textwidth]{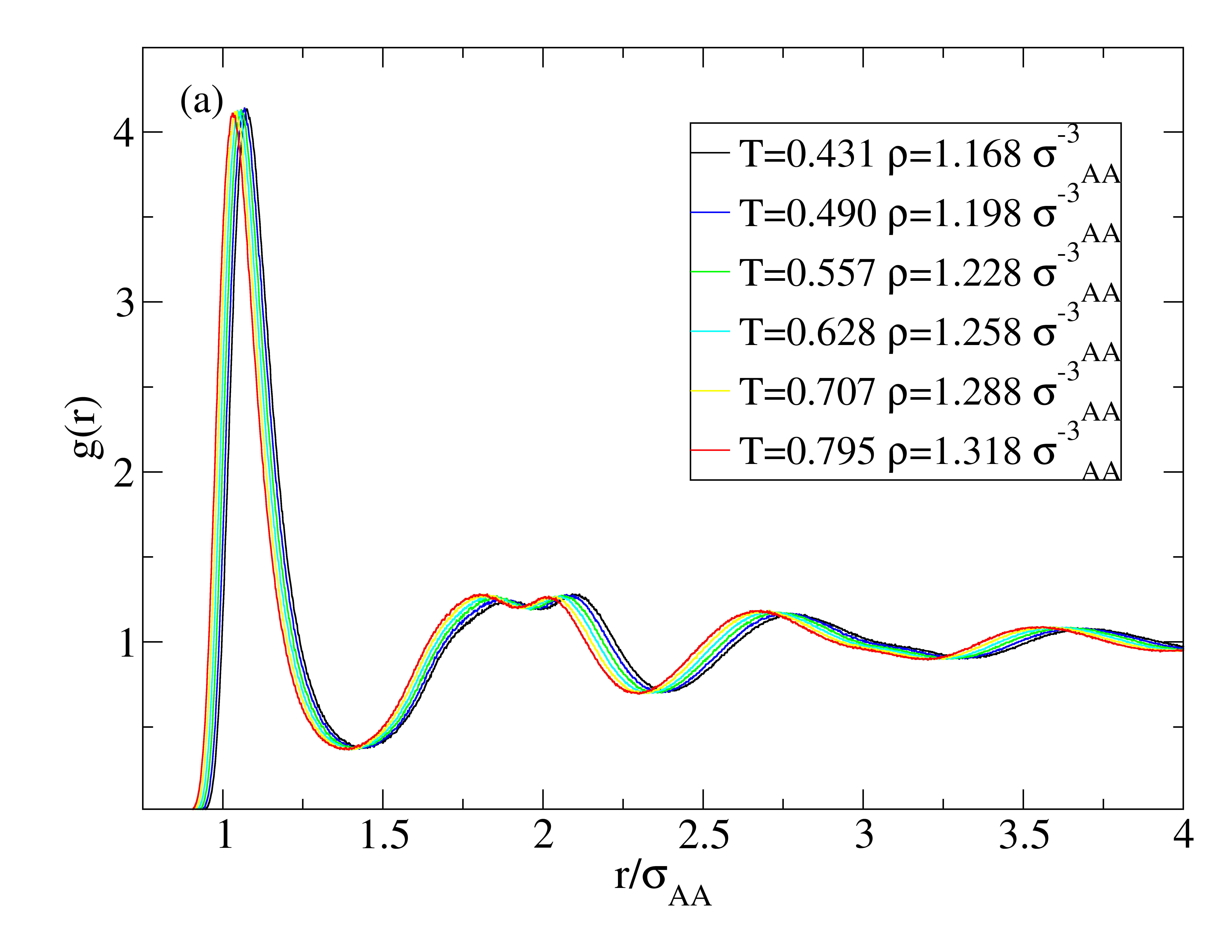}
\includegraphics[width=0.3\textwidth]{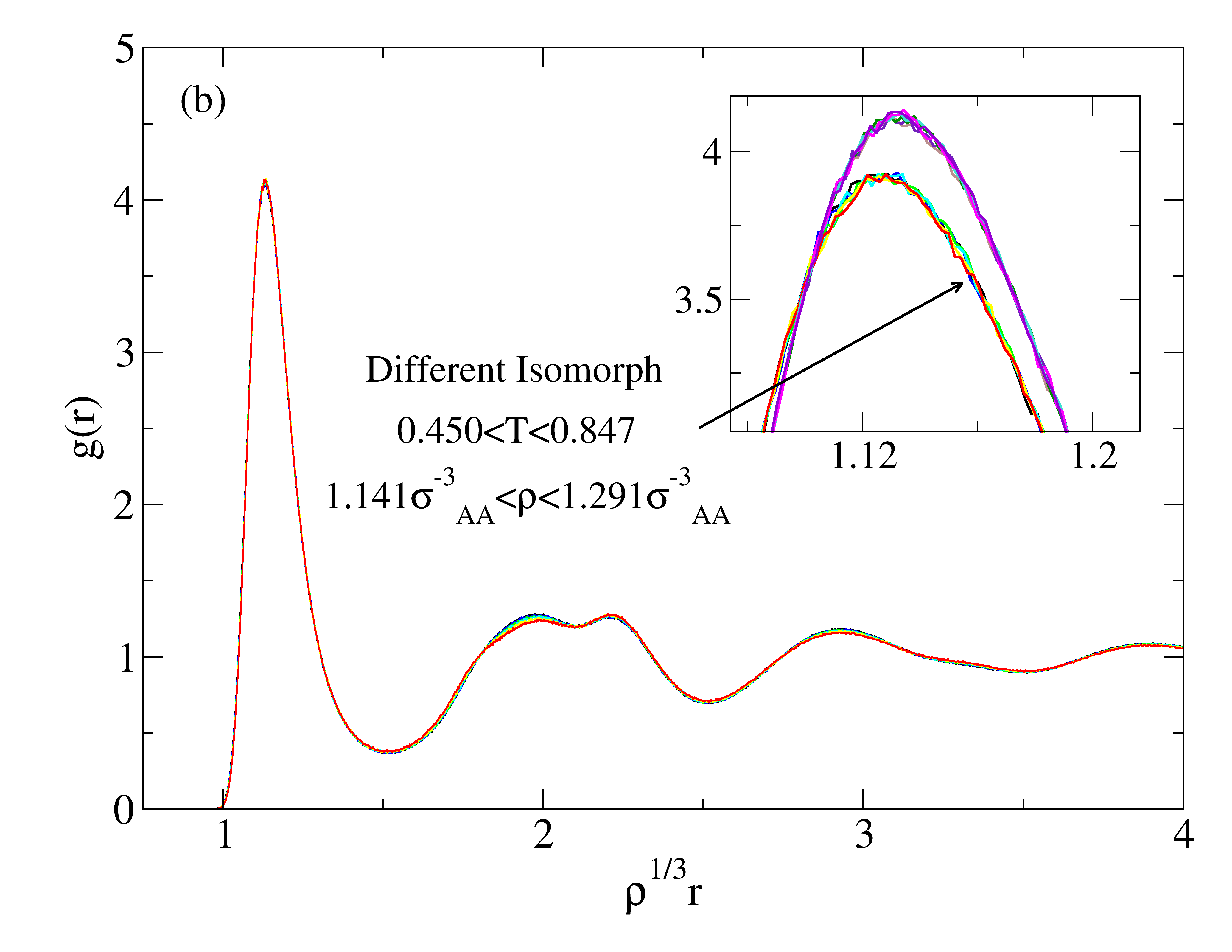}
\includegraphics[width=0.3\textwidth]{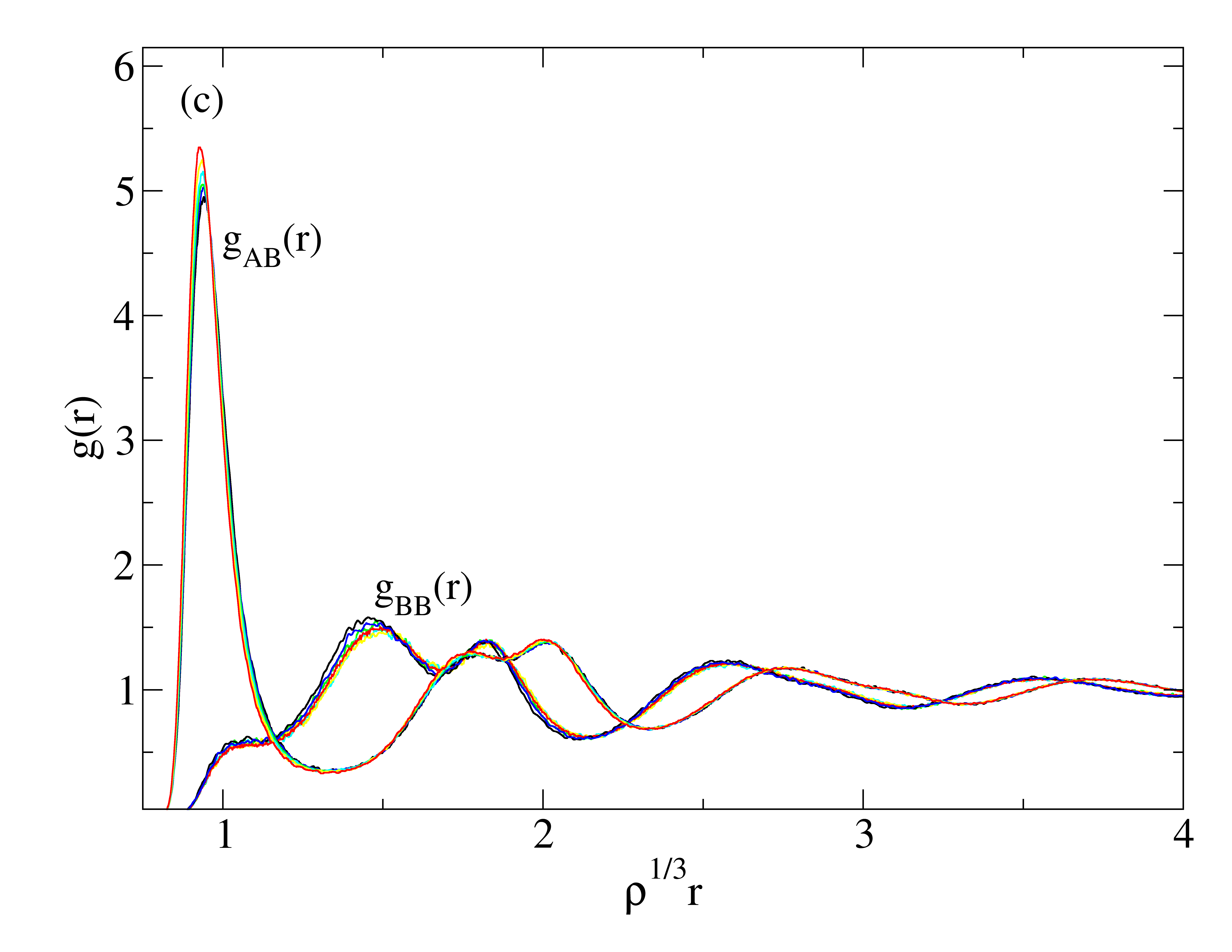}
\end{center}
\caption{(a) AA particle radial distribution functions for the KABLJ liquid at six isomorphic state points. 
(b) The same radial distribution functions plotted as functions of the reduced distance, showing very good data collapse. The inset focuses on the first peak and includes results for a second collection of isomorphic state points.
(c) AB and BB radial distribution functions in reduced units for the six isomorphic state points of (a). There is good data collapse, but with larger deviations than for the AA distribution function. This shows that isomorph properties are not exact.}\label{fig:figure2}
\end{figure} 

Denoting the large Lennard-Jones particle as A,  Fig. \ref{fig:figure2}(a) gives the AA radial distribution functions for six isomorphic state points of the KABLJ system. The temperature varies by almost a factor of two; nevertheless there is good collapse of the curves when these are plotted as functions of the reduced distance (Fig. \ref{fig:figure2}(b)). The inset of (b) zooms in on the peak of the radial distribution function. We added here data for a second collection of isomorphic state points, showing that different isomorphs have different structure. That the isomorph invariants are, after all, not exact is clear from Fig. \ref{fig:figure2}(c), which shows the scaled AB and BB radial distribution functions of the isomorphic state points of (a).

\begin{figure}
\begin{center}
\includegraphics[width=0.4\textwidth]{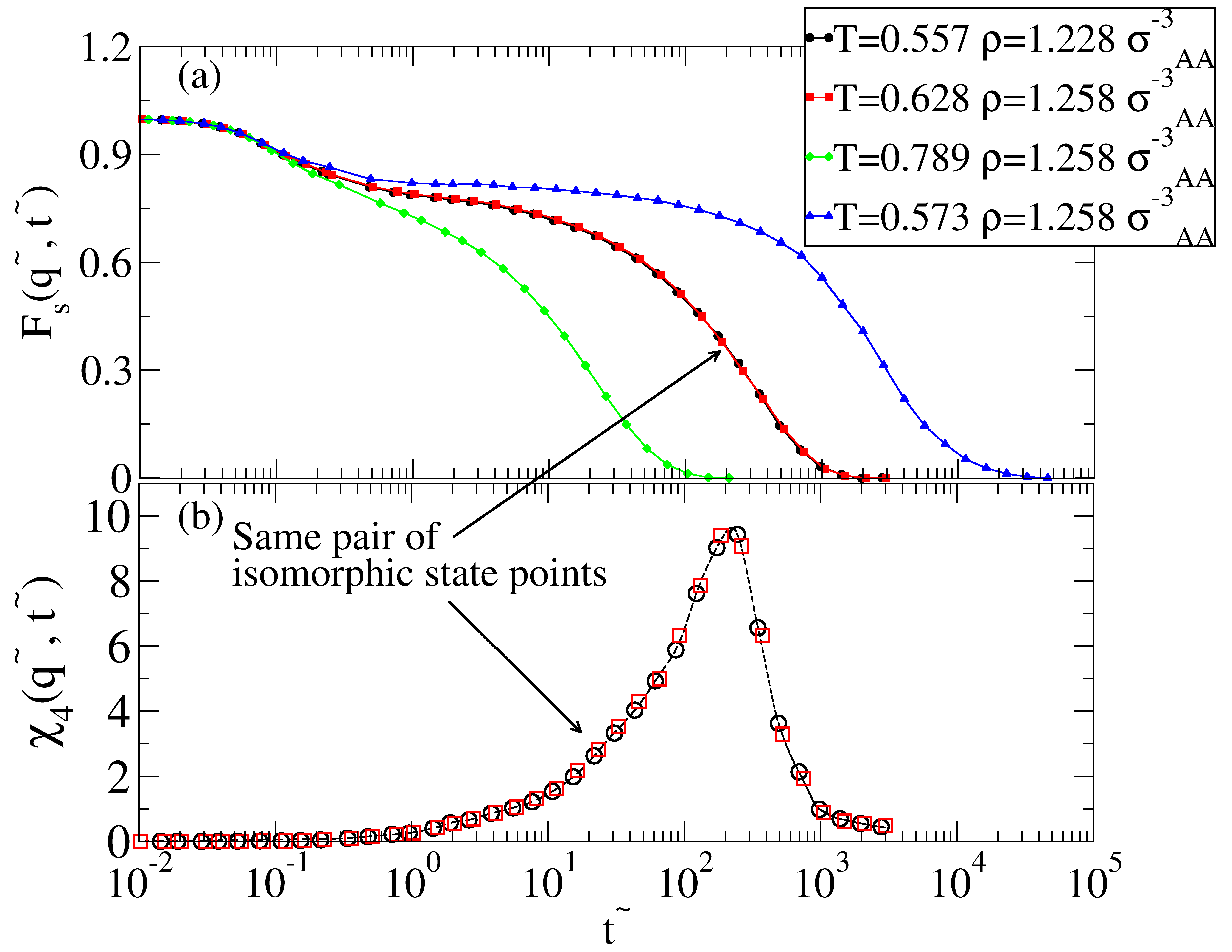}
\end{center}
\caption{\footnotesize{(a) The self part of the intermediate scattering functions at the wavevector corresponding to the first peak of the static structure factor as function of the reduced time $\tilde t$ for four state points of the KABLJ liquid. Two of the state points are isomorphic (red and black curves); these two state points have virtually indistinguishable self part of the intermediate scattering functions.\\
(b) The four-point dynamic susceptibility $\chi_4(\tilde{t})$ for the two isomorphic state points of (a). This quantity, which measures the degree of dynamic heterogeneity, is predicted to be an isomorph invariant. The full curve is a cubic spline fit to the black data points.}} \label{fig:figure1}
\end{figure}

Turning now to the dynamics, Fig. \ref{fig:figure1}(a)  shows the AA self part of the intermediate scattering functions $F_s(\tilde q,\tilde t)$ at the reduced wavevector corresponding to the first peak of the static structure factor, where $\tilde t$ is time scaled by the characteristic time $t_0=\rho^{-1/3}(m/k_BT)^{1/2}$ (Table \ref{red_table}). The figure shows results for four state points at two densities. Two of the four state points are isomorphic. These two state points (black and red curves) have virtually identical relaxation behavior, including the short-time ``cage-rattling''. Figure \ref{fig:figure1} confirms isomorph properties 3c and 3d. For the two isomorphic state points of Fig. \ref{fig:figure1}(a), (b) shows the four-point dynamic susceptibility defined\cite{ton05} by $\chi_4(\tilde t)=N\langle (\Delta F_s(\tilde q,\tilde t))^2\rangle$, i.e., the mean-square fluctuation of the self-part of the intermediate scattering functions. The quantity $\chi_4(\tilde t)$  measures the degree of dynamic heterogeneity on a given time scale. By the predicted isomorph invariance of time-correlation functions as functions of reduced time (property 3c) $\chi_4(\tilde t)$ is predicted to be invariant, which is indeed the case.\cite{cos09b,fra09}

\begin{figure} 
\begin{center}
\includegraphics[width=0.4\textwidth]{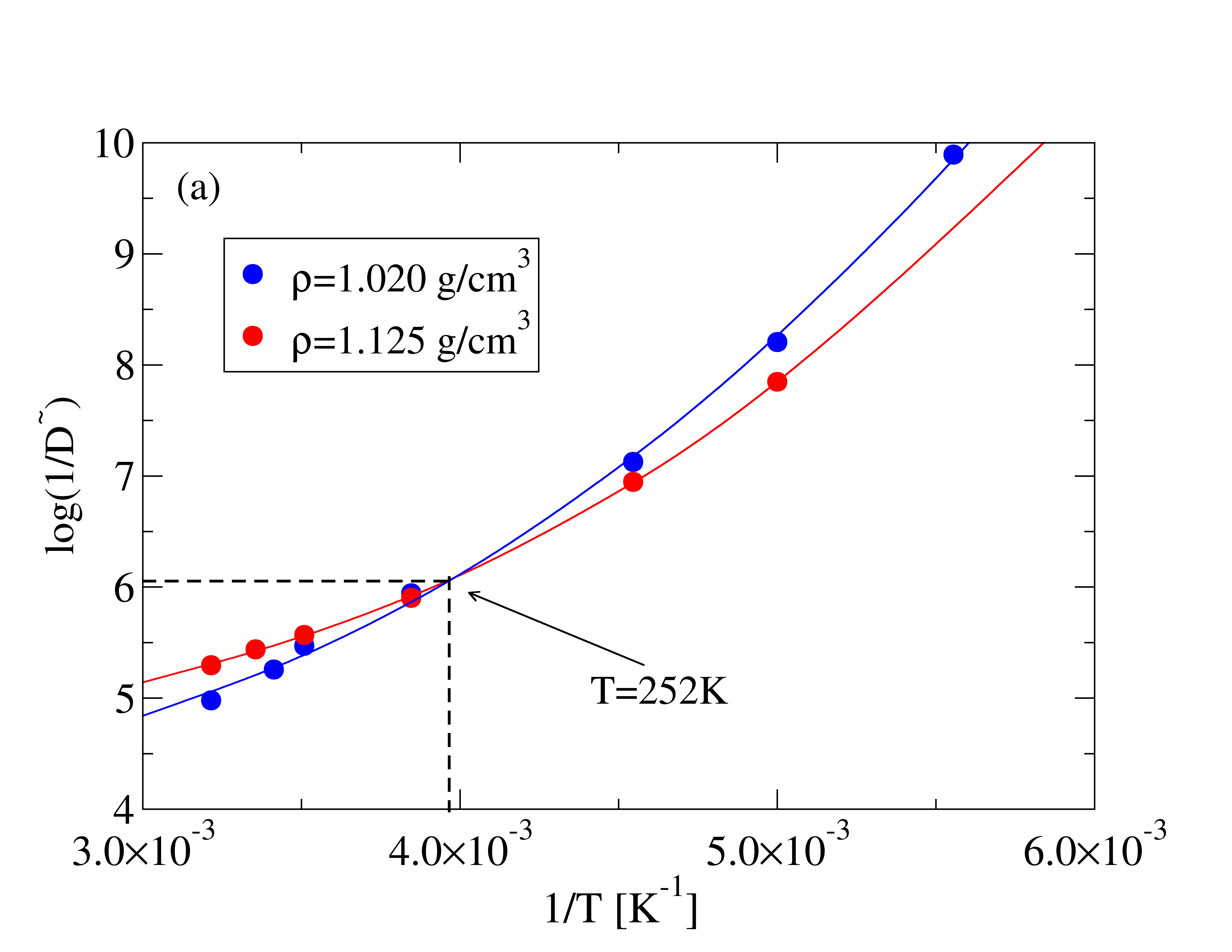}
\includegraphics[width=0.4\textwidth]{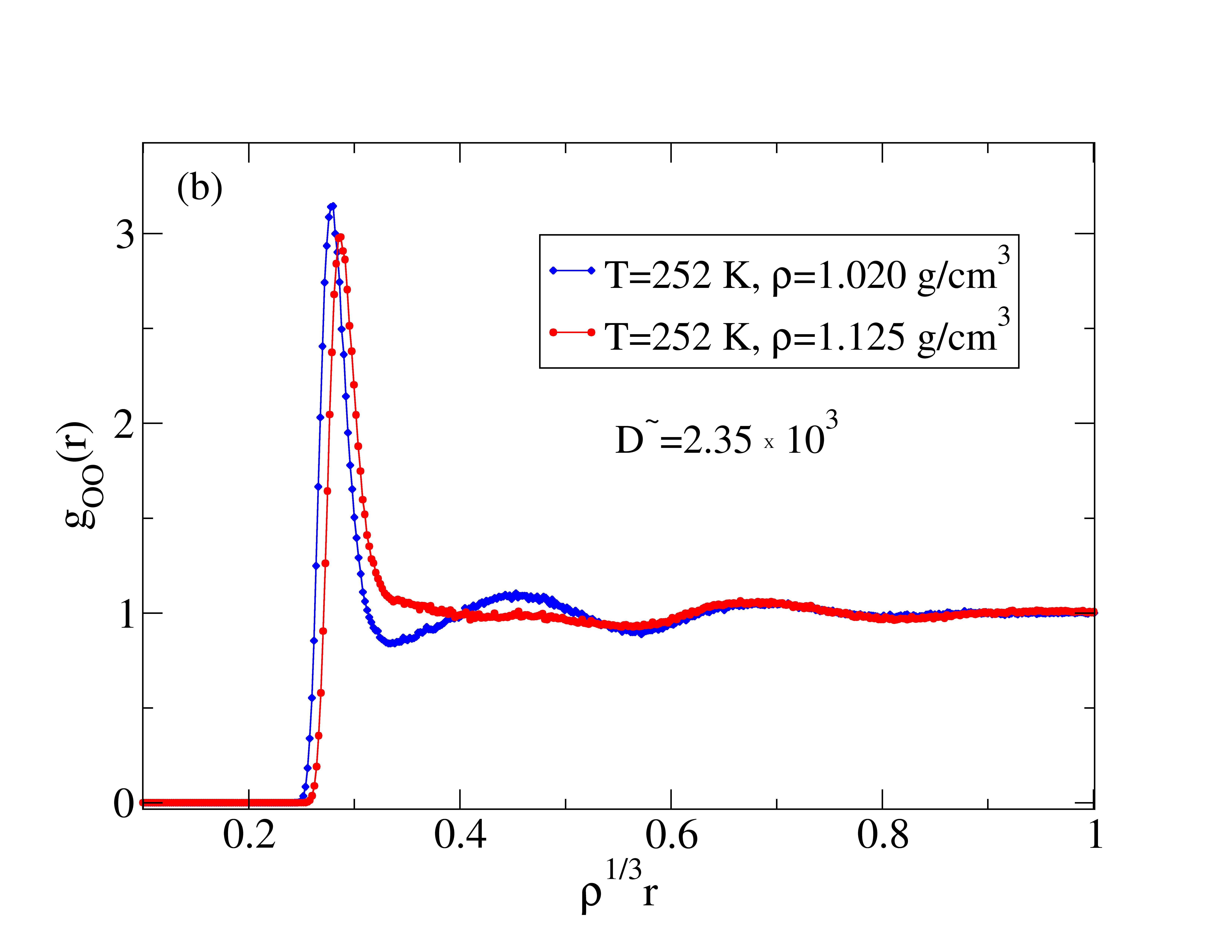}
\end{center}
\caption{
(a) Reduced diffusion constants for two sets of isochoric state points of the SPC water model, which is not strongly correlating. The data were fitted by a polynomial in order to identify the temperature where the two densities have the same reduced diffusion constant (T=252K).\\
(b) Oxygen-oxygen radial distribution functions as functions of reduced distance for the two state points of SPC water identified in (a) with same temperature and same reduced diffusion constant, but different density. If water had isomorphs, the two radial distribution functions would be virtually identical.}\label{fig:figure3}
\end{figure}

As shown briefly in Sec. II and in detail in the Appendix A, only strongly correlating liquids have isomorphs. A system that is not strongly correlating\cite{I} is the SPC water model, where the hydrogen bonds are mimicked using Coulomb interactions and the oxygen atoms interact via an LJ potential.\cite{SPC} The fact that this model has near zero $WU$ correlation reflects\cite{I} water's density maximum. In order to prove that SPC water does not have isomorphs, suppose that it did. Then state points with same reduced diffusion constant would have same structure. Figure \ref{fig:figure3}(a) shows the reduced diffusion constant as a function of temperature for two sets of isochoric state points. Interpolation with a polynomial was done in order to identify the temperature where the two densities have the same reduced diffusion constant (T=252 K). Figure \ref{fig:figure3}(b) shows the radial distribution functions of these two state points. Clearly, SPC water does not have isomorphs.

\subsection{Out-of-equilibrium properties: Aging}

We showed by example that isomorphic state points have the same scaled static and dynamic correlation functions, but all properties tested so far were equilibrium properties. What happens when a strongly correlating liquid is taken out of equilibrium? To answer to this we simulated temperature/density jumps from equilibrium, i.e., instantaneous changes of these two variables to new values. All states involved in the ``aging experiments'' belong to the two sets of isochoric points whose self-intermediate scattering functions and radial distribution functions are plotted in Figs. \ref{fig:figure2}(a) and \ref{fig:figure2}(b).

\begin{figure}
\begin{center}
\includegraphics[width=0.4\textwidth]{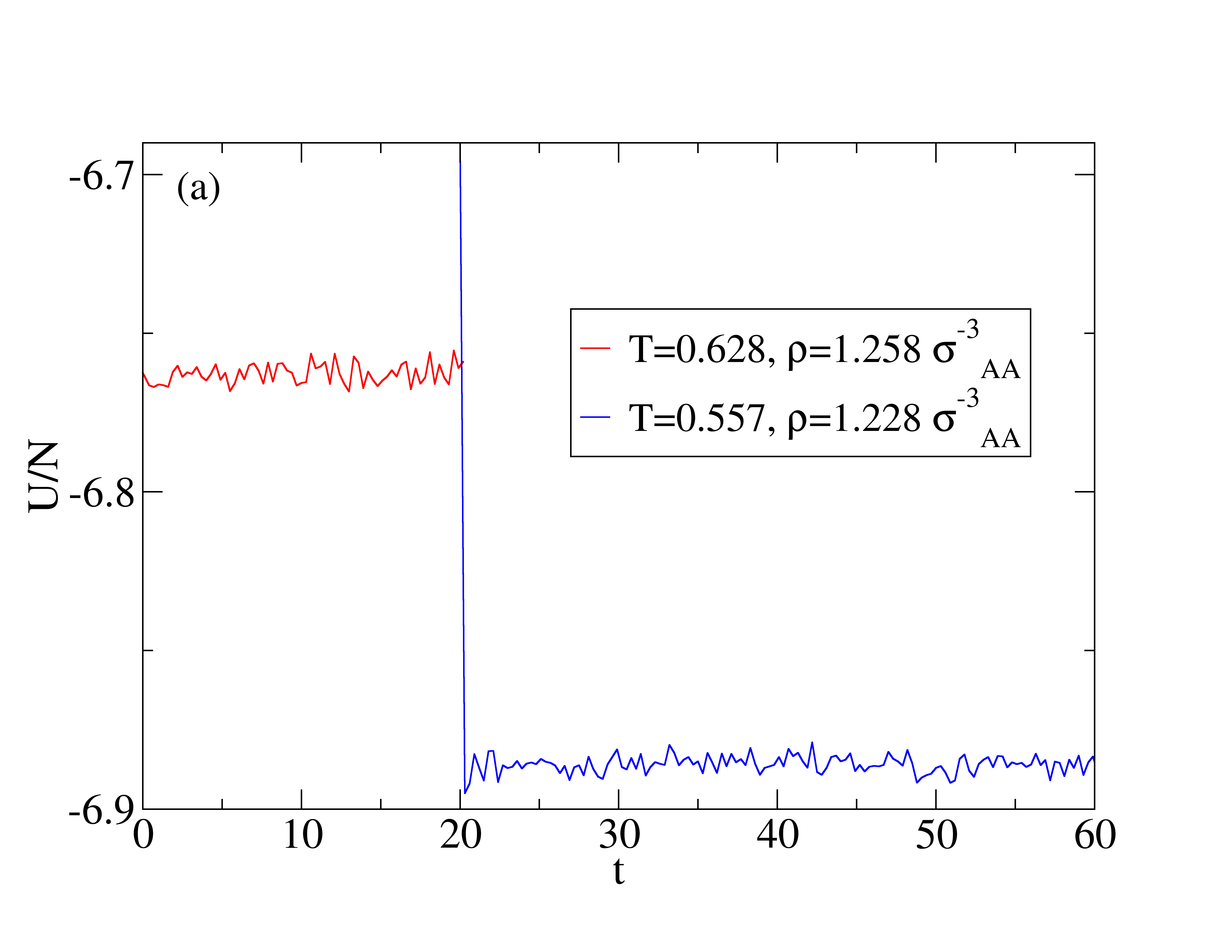}
\includegraphics[width=0.4\textwidth]{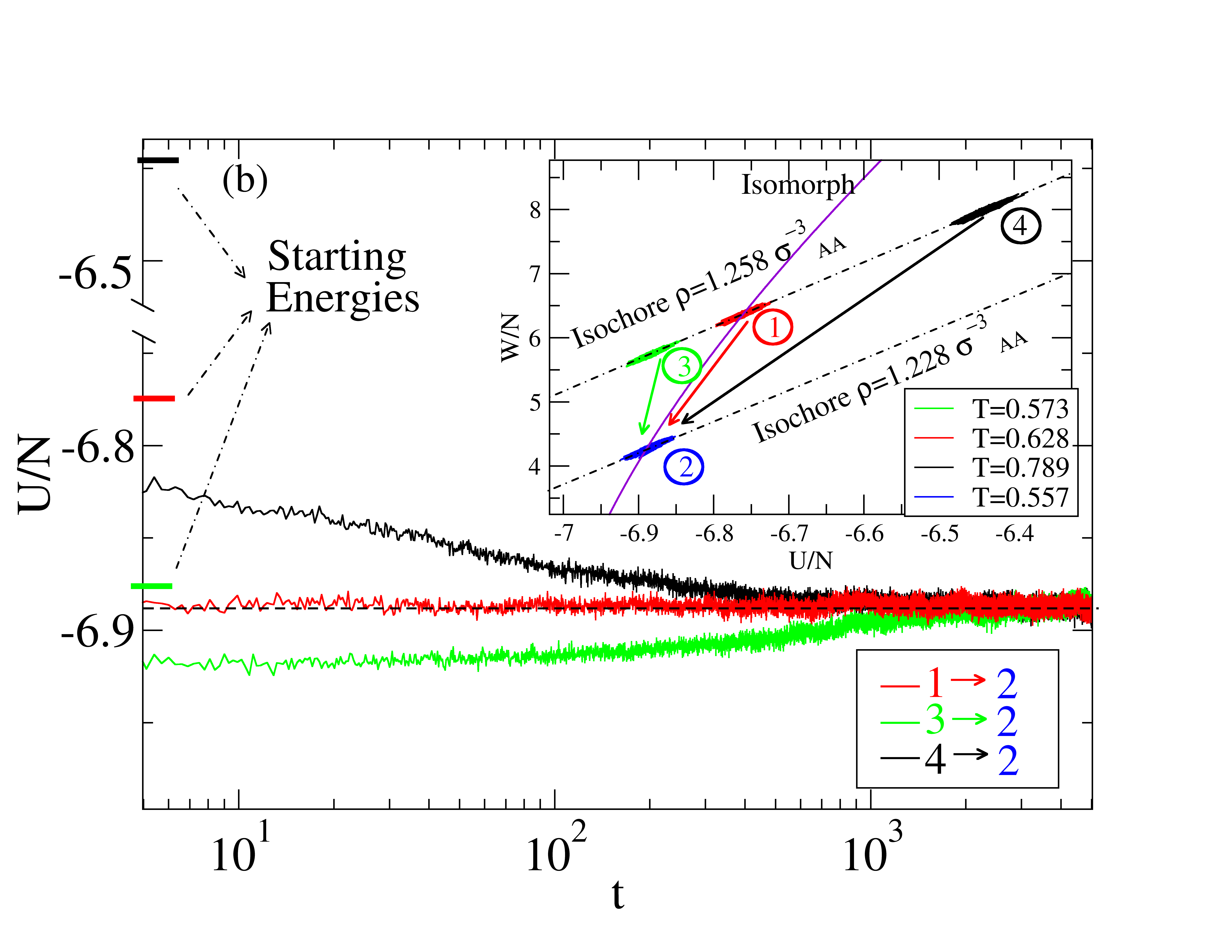}
\end{center}
\caption{(a) Results from simulating an instantaneous temperature and density jump applied to the KABLJ liquid, starting from equilibrium and jumping to a state which is isomorphic with the initial state (LJ units). Except for the transient no relaxation is associated with the jump, showing that the system is instantaneously in equilibrium. The relaxation time of both state points is about $500$ in reduced units (the two state points have the same reduced relaxation time because they are isomorphic).\\
(b) Potential energy relaxation towards equilibrium for the KABLJ liquid for different jumps: $1\rightarrow 2$, $3\rightarrow 2$, and $4\rightarrow 2$ (see the inset). The $1\rightarrow 2$ jump is the isomorphic jump of (a), the two other jumps are not between isomorphic state points. 
-- The results shown in (a) and (b) are averaged over ten independent simulations.}\label{fig:figure4}
\end{figure}

Figure \ref{fig:figure4}(a) shows the time evolution of the potential energy when a jump is made from equilibrium, bringing the KABLJ liquid to a new state point which is isomorphic to the initial state. For both state points the reduced relaxation time is around 500. The jump was performed as follows: We instantaneously increased the box volume without changing particle positions (the initial overshoot is due to this) and simultaneously changed the thermostat temperature to the final temperature. There are no signs of slow relaxation after the jump. Thus the system is immediately in equilibrium, as predicted for jumps between isomorphic state points (property 4a). 

The horizontal (red) line in Fig. \ref{fig:figure4}(b) shows the data of Fig. \ref{fig:figure4}(a), now on a logarithmic scale with time shifted such that the jump  occurs at $t=0$. Here it is even more clear that jumps between two isomorphic state points preserve equilibrium. In contrast, the $3\rightarrow 2$ and $4\rightarrow 2$ jumps both age slowly to equilibrium, where all three initial states (1), (3), and (4) have the same density. Clearly, instantaneous equilibration is a feature of jumps between isomorphic points only.

\begin{figure}
\begin{center}
\includegraphics[width=0.5\textwidth]{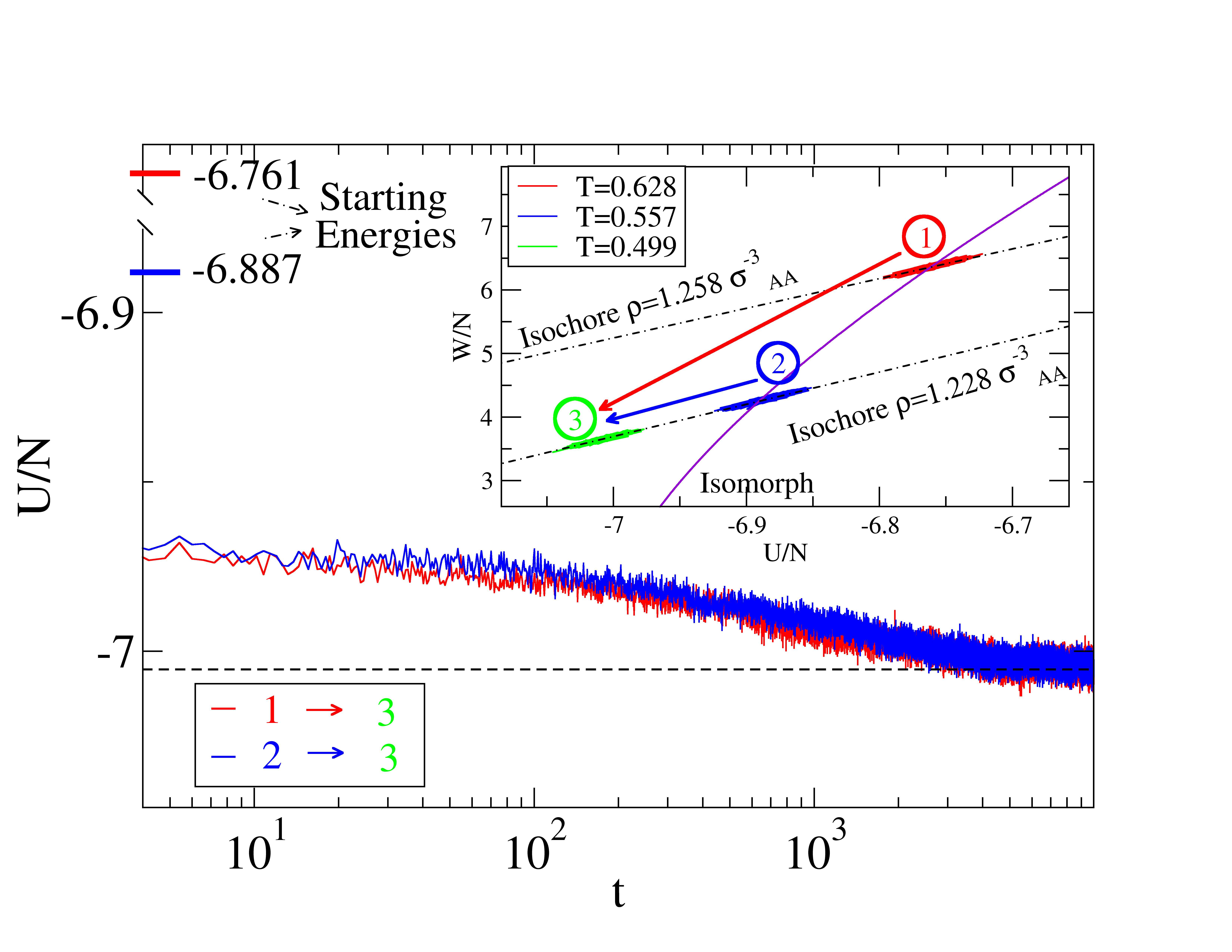}
\end{center}
\caption{Potential energy relaxation towards equilibrium for the KABLJ liquid comparing the jumps $1\rightarrow 3$ and $2\rightarrow 3$ (see the inset). The two relaxations are predicted to be identical because state points (1) and (2) are isomorphic. The results shown are averaged over ten independent simulations.}\label{fig:figure5}
\end{figure} 

The preservation of equilibrium for jumps between isomorphic state points has important consequences for general jumps (Fig. \ref{fig:figure5}). Consider first the inset of Fig. \ref{fig:figure5}. Suppose we start in equilibrium at state point (1) and change temperature and density to $T_3$ and $\rho_3$. According to property 4b, if state point (2) is isomorphic to state point (1) the observed relaxation behavior for the $2\rightarrow 3$ jump should be the same as for the $1\rightarrow 3$ jump. The simulations confirm this prediction. 

We can now also understand why the $3\rightarrow 2$ jump in Fig. \ref{fig:figure4}(b) approaches the equilibrium potential energy from below, although state point (3) has an average potential energy which is slightly {\it higher} than that of state point (2) (compare the inset of Fig. \ref{fig:figure4}(b)). This is because, whenever a jump is performed, the system makes an instantaneous isomorphic jump to the new density to relax from here to the final state point by moving on the isochore. Thus in Fig. \ref{fig:figure4}(b), the $3\rightarrow 2$ jump is first an isomorphic jump to the state with correct density, a state that has lower average potential energy than state (2). Only thereafter the system relaxes towards state (2).

In most experiments pressure, not density, is controlled. The only difference is that an instantaneous isomorphic jump goes to a new pressure, instead of to a new density, and that the subsequent relaxation follows an isobar instead of an isochore. Thus relabeling the dashed lines in the insets  of Figs. \ref{fig:figure4}(b) and \ref{fig:figure5} to isobars would qualitatively illustrate an experimental temperature and pressure jump.

\section{Relating isomorphs to some topics of current liquid state theory and experiment}

This section discusses connections between the isomorph concept and current liquid state theory and experiment. No subject is treated in depth; the purpose is merely to show by example that isomorphs fit nicely into a number of previous findings. In most cases we demonstrate how these are {\it consistent} with the isomorph concept, but in a few cases a previous finding is shown to be a {\it consequence} of the existence of isomorphs for strongly correlating liquids.

\subsection{Phenomenological melting rules} 

Melting is an old subject of condensed-matter physics. There has long been a good understanding of the statistical mechanics of melting via the density functional theory of Ramakrishnan and Yussouff.\cite{ram79} Supplementing this are a number of phenomenological melting rules. Perhaps the most famous is the Lindemann criterion, according to which melting takes place when the crystal's vibrational root mean-square displacement is about 10\% of the nearest-neighbor distance.\cite{gil56} 

For a strongly correlating liquid the melting curve in the phase diagram must be parallel to nearby liquid and crystalline isomorphs (the isomorph concept applies to the crystalline phase as well as to the liquid). This is because, if an isomorph were to cross the melting curve, the Boltzmann factors favoring crystalline order would dominate one part of that isomorph and be negligible on another part; this contradicts the structure invariance. For strongly correlating liquids the invariance of the crystalline excess entropy along the melting line implies, as is easy to show, pressure invariance of the Lindemann melting criterion. Moreover, since both crystalline and liquid excess entropy are invariant along the melting curve, the melting entropy must be pressure independent. 

There are other consequences of melting curves being isomorphs for strongly correlating liquids. Thus along the melting curve, slightly to the liquid side, a number of properties are predicted to be invariant: the reduced viscosity, the reduced surface tension, the reduced diffusion constant, the reduced heat conductivity -- in fact all isomorph invariants. There are several theoretical and experimental works pointing in these directions.\cite{hoo71,ros75,sti75,poi88,rut02,dig04} For instance, the reduced-unit static structure factor of liquid iron measured by x-ray scattering is invariant along the melting curve for pressures up to 58 GPa.\cite{she04} Finally, note that the Hansen-Verlet criterion,\cite{han69} according to which melting takes place when the liquid structure factor peak is $2.85$, due to the invariance of the structure factor in reduced coordinates is also consistent with the melting curve being an isomorph for strongly correlating liquids.

\subsection{Rosenfeld's excess entropy scaling} 

Rosenfeld presented in 1977 an interesting observation:\cite{ros77,ros99} For a large class of model systems the reduced transport coefficients (table I) appear to be functions of the excess entropy only. He justified what is now known as ``excess-entropy scaling'' by reference to variational hard-sphere thermodynamic perturbation theory.\cite{ros77} He emphasized that excess-entropy scaling is a semi-quantitative model rather than a theory and referred to excess-entropy scaling as a principle of corresponding states. For any given strongly correlating liquid excess-entropy scaling follows from the isomorph properties derived in Sec. II, although this does not explain why the functional excess entropy dependence appears to be quasiuniversal.\cite{ros77,ros99} Recent simulations of the Gaussian core model by Truskett and co-workers\cite{kre09} provide an example where Rosenfeld's excess-entropy scaling fails. This is not surprising, since this model due to the soft core is not expected to be strongly correlating.  

A scaling procedure that is similar in spirit to Rosenfeld's excess entropy scaling was discussed by Dzugutov in 1996.\cite{dzu96} He showed from simulations that the reduced diffusion constant follows $\tilde D\propto\exp(S_2/Nk_B)$ for a number of systems, where $S_2$ is the two-particle entropy. This relation is also consistent with the isomorph concept, because both $\tilde D$ and $S_2$ are both isomorph invariants (properties 3e and 2b). 

From the isomorph viewpoint there is no reason to expect Rosenfeld's or Dzugutov's equations to hold for liquids that are not strongly correlating. While genuine Rosenfeld scaling seems to fail for liquids that are not strongly correlating, it appears that related excess entropy scaling procedures hold for many such liquids (see, e.g., Refs. \onlinecite{kre09a} and \onlinecite{pon09}). Why this is, is an important question for future research.

\subsection{Young and Andersen's approximate scaling principle}

In 2003 Young and Andersen conjectured an approximate scaling principle according to which two liquids have similar dynamical properties if they at two same-density state points have similar static pair-correlation functions -- even if their potentials are quite different and the temperatures differ.\cite{you03,you05} This was confirmed by simulations comparing the LJ liquid to the same system with a Weeks-Chandler-Andersen cut-off where only the repulsive part of the potential is kept. At same density the latter system required lower temperature in order to have almost identical pair-correlation functions. When temperature was properly adjusted, however, the two liquids were shown to have similar coherent and incoherent intermediate scattering functions, as well as similar velocity and current auto-correlation functions. This confirms the proposed approximate scaling principle for the single-component LJ liquid (in contrast, Berthier and Tarjus recently showed that the WCA approximation applied to the KABLJ liquid does not reproduce this liquid's dynamics\cite{ber09}). 

As pointed out by Young and Andersen, if not only the pair- but all higher-order correlation functions are identical for two liquids, they must have proportional NVT Boltzmann statistical weights. This suggests generalizing the isomorph concept to an equivalence relation between {\it different} liquids. One may define two liquids to be isomorphic if: 1) They are both strongly correlating, and 2) a pair of state points exists such that Eq. (\ref{isodef}) is obeyed for all physically relevant microscopic configurations with the same reduced coordinates in the form generalized to two different potential energy functions, $U_1$ and $U_2$. Clearly when this is obeyed, the entire system of isomorphic curves of one liquid maps onto that of the other liquid.

\subsection{Two-order parameter maps of Debenedetti and coworkers}

In a series of publications Debenedetti and collaborators studied liquid structure in terms of a translational order parameter, $t$, and a system-dependent orientational order parameter that may be, for instance, the often used quantity $Q_6$.\cite{tor00,tru00,err01,she02,err03,gio05} The translational order parameter is defined as $t=\int_0^{{\tilde r}_c} |g(\tilde r)-1|d \tilde r$ where $g(\tilde{r})$ is the reduced-coordinate pair distribution function and $\tilde r_c =\rho^{1/3}r_c$ is a cut-off; $Q_6$ is defined from sixth-order spherical harmonics involving the ``bonds'' to the twelve particles nearest a given particle (averaged over all particles). The general picture one finds by plotting $t$ versus $Q_6$ in so-called ordering maps is that for the LJ liquid and related systems there is a striking collapse, showing that ``bond-orientational order and translational order are not independent for simple spherically symmetric systems at equilibrium''.\cite{err03} For water\cite{err01,gio05} and silica,\cite{she02} on the other hand, there is no such collapse; here the order parameters cover a two-dimensional region of order-parameter space. These results may be explained by reference to isomorphs: Since both order parameters are isomorph invariants, they cannot vary independently for a strongly correlating liquid. Water and silica are not strongly correlating (Paper I), so there is no reason to expect their two order parameters to follow each other.

\subsection{Viscous liquid dynamics}

The dynamics of viscous liquids is a long-time focus of our research at the DNRF {``Glass and Time''} center, and the class of strongly correlating liquids was identified in our efforts to understand the properties of single-parameter viscous liquids.\cite{ped08a,ell07,bai08c} As shown below, the isomorph concept throws new light on facts and puzzles of the physics of viscous liquids.

1) {\it Cause of the non-Arrhenius relaxation time: The ``isomorph filter''.}
A major puzzle concerning viscous liquids approaching the glass transition is the origin of the non-Arrhenius temperature dependence of their average relaxation time $\tau$.\cite{kau48,bra85,gut95,edi96,cha98,ang00,alb01,deb01,bin05,sci05,dyr06,sch86,hod94,avr96} In many cases the relaxation-time increase is truly dramatic, with a factor of ten or more slowing down if temperature is decreased by just 1\%. There are several models for this. Although it is not obvious that any simple universal model exists, most workers in the field assumes this to be the case. Any universal model must apply to strongly correlating liquids, of course. This implies a criterion which we shall refer to as the isomorph filter. The idea is the following. Since the reduced average relaxation time is an isomorph invariant, this quantity can only be controlled by a quantity that is also an isomorph invariant. All theories relate $\ln\tau$ to some quantity. For any viscous liquid $\ln\tau\cong\ln\tilde\tau$, because the temperature variation of the average relaxation time dominates completely over the reduced time unit's $\sqrt T$ temperature dependence. Thus the average relaxation time must be controlled by an isomorph invariant. Consider now some examples.

According to the Adam-Gibbs model\cite{ada65,dyr09} the relaxation time $\tau$ varies as $\ln\tau\propto 1/TS_ {\rm conf}(\rho,T)$, where $S_ {\rm conf}(\rho,T)$ is the configurational entropy discussed in connection with isomorph property 1b. Adam and Gibbs wrote the proportionality constant as the product of a critical configurational entropy $s_c^{\ast}$ and a term $\Delta\mu$ which is ``largely the potential energy hindering the cooperative rearrangement per monomer segment'', and they argued that the state-point dependence of these terms can be neglected. This is also what is usually done in comparing to experiments. In this case, however, since $\tilde\tau$ and $S_ {\rm conf}(\rho,T)$ are both invariant along an isomorph, the model is inconsistent with the isomorph invariance due the extra factor $T$. Thus in this version the model cannot apply to strongly correlating liquids. Alternatively, the proportionality constant must be allowed to vary with state point. Indeed, simulations of Sastry\cite{sas00} of the KABLJ liquid show good agreement with the Adam-Gibbs model if the constant is density dependent, a density dependence which, in order not to violate isomorph invariance, must be -- and to a reasonable approximation is -- given by a factor proportional to $\rho^\gamma$. Although the original Adam-Gibbs model cannot apply for strongly correlating liquids (van der Waals liquids and metallic liquids), it should be noted that for very fragile liquids the temperature dependence of $S_ {\rm conf}(\rho,T)$ dominates over the $T$ factor in the Adam-Gibbs expression;\cite{ada65} for these liquids the model may work fairly well. 
Another possibility is that $S_ {\rm conf}(\rho,T)$ alone controls the relaxation time. Interestingly, this was suggested by Bestul and Chang in a paper\cite{bes64} preceding Adam-Gibbs by one year, which showed data consistent with a universal value of the excess entropy at the glass transition for several liquids. 
Much more recently, Truskett and collaborators\cite{mit06} suggested that the excess entropy $\sex$ controls the relaxation time in the spirit of Rosenfeld scaling;\cite{ros77,ros99} this is also consistent with isomorph invariance.
As a final example of a theory where entropy controls the relaxation time, consider the random first-order transition (RFOT) of Wolynes and co-workers (reviewed in  Ref. \onlinecite{lub07}). According this theory and its generalization by Bouchaud and Biroli,\cite{bou04} for some exponents $x$ and $y$ the relaxation time is given $\ln\tau\propto (Y/T)^xS_ {\rm conf}^{-y}(\rho,T)$ where $Y$ is a surface tension at the molecular scale. This expression is isomorph invariant only if $Y\propto T$. This is, however, precisely what was predicted in 2000 on different grounds by Xia and Wolynes,\cite{xia00} so the RFOT passes the isomorph filter.

The free-volume model of Cohen and Grest predicts that $\ln\tau\propto 1/v_f(\rho,T)$ where $v_f(\rho,T)$ is the free volume. As it stands, this model does not survive the isomorph filter because the proportionality constant is state-point independent and, although the free volume definition is not obvious, $v_f(\rho,T)$ is hardly an isomorph invariant. If one imagines $v_f$ to be a geometrically determined quantity measured in units of $1/\rho$, however, the model {\it is} invariant along an isomorph. Thus, while $v_f(\rho,T)$ cannot determine the relaxation time, the quantity $v_f(\rho,T)\rho$ can possibly. 
The shoving model\cite{dyr06,dyr96,tor09} predicts that $\ln\tau\propto G_\infty(\rho,T)V_c/k_BT$, where $V_c$ is a characteristic volume which in experiments is of order the molecular volume. This model, which is one of several related ``elastic'' models,\cite{dyr06} is consistent with the existence of isomorphs if $V_c$ is geometrically determined:  In this case $V_c\propto 1/\rho$ and, because $G_\infty(\rho,T)/T\rho$ is an isomorph invariant (property 3f), this implies that $G_\infty(\rho,T)V_c/k_BT$ is isomorph invariant. 
The vibrational mean-square displacement version of the elastic models, $\ln\tau\propto a^2/\langle x^2\rangle$, is not isomorph invariant if $a$ is assumed to be independent of density,\cite{lep08} but it does survive the isomorph filter if the reasonable assumption is made\cite{nis09} that $a\propto\rho^{-1/3}$.
As a final example, we note that the entropic barrier hopping theory of Schweizer and coworkers\cite{sch07,tri09} in a certain limit predicts that $\ln\tau\propto d^2\langle F^2\rangle /(k_BT)^2$ where $d$ is an effective hard sphere diameter and $\langle F^2\rangle$ is the short-time-averaged single-particle mean-square force. This is expression is not isomorph invariant, but the modest modification of it obtained by replacing $d$ by $\rho^{-1/3}$ leads to an isomorph invariant expression, as is easily shown by use of Eq. (\ref{uinv}).

2) {\it Isochronal superposition.}
Building on earlier works by Roland {\it et al.},\cite{rol03} in 2005 Ngai, Casalini, Capaccioli, Paluch, and Roland published a paper entitled ``Do theories of the glass transition, in which the structural relaxation time does not define the dispersion of the structural relaxation, need revision?''\cite{nga05} The authors showed here that for many viscous liquids and polymeric systems the average relaxation time determines the shape of the dielectric loss peak. Thus whether the average relaxation time is increased by lowering temperature or by increasing pressure, the effect is the same on the relaxation time distribution as monitored via the dielectric loss. This result, which we for brevity refer to as ``isochronal superposition'', was puzzling at the time (at least to us). Isochronal superposition now appears as a consequence of the existence and properties of isomorphs: Any strongly correlating liquid must obey isochronal superposition because, according to properties 3c and 3d, both the average relaxation time and the dielectric spectrum are isomorph invariants (when given in reduced units, but as mentioned their use makes little difference for viscous liquids). Thus whether temperature is lowered or pressure is increased (or a combination thereof) to bring the liquid to state points with the same average relaxation time, all such state points are isomorphic. Consequently, they have the same relaxation spectra for quantities probed by linear response experiments. In Ref. \onlinecite{nga05} hydrogen-bonding liquids were quoted explicitly as exceptions to isochronal superposition; this is consistent with the fact that these liquids are not strongly correlating.\cite{II}

3) {\it How many parameters are needed for describing a viscous liquid?} In the old days of glass science it was actively debated whether one or more ``order parameters'' are required to describe glass structure and the glass transition. We write ``order parameters'' in quotation marks in order to emphasize that this term has a somewhat different use in traditional glass science than in the theory of critical phenomena. The original considerations of Prigogine, Defay, Davies, and Jones, and others, referred to the glass transition as a second-order phase transition in the Ehrenfest sense.\cite{pri54,dav52,dav53} It was recently shown that strongly correlating liquids are precisely the liquids that to a good approximation may be regarded as single-order-parameter liquids;\cite{ped08a,ell07} a review of the connection was given in Ref. \onlinecite{bai08c}. Strongly correlating liquids have isomorphs, so it should be possible to link isomorphs directly to the single-order-parameter scenario. Indeed, by labelling the collection of isomorphic curves in the phase diagram with a continuously varying real number, a formal single-order-parameter description throughout the phase diagram is arrived at. This suggests a generic way of defining the single-order-parameter scenario that is implicit in the old works. The isomorph label is not unique, obviously; it may be chosen as any property that is isomorph invariant, for instance $\ccv$.

\subsection{Some further potential isomorph connections}

1) In 1989 Baranya and Evans\cite{bar89} reported from simulations that the excess entropy of the LJ liquid to a good approximation equals the two-particle entropy $S_2$ plus a constant. This is consistent with the existence of isomorphs, because on an isomorph both $S_2$ and $S_ {\rm ex}$ are invariant. That is, because the LJ liquid is strongly correlating, if for instance $S_ {\rm ex}$ were an isomorph invariant but $S_2$ were not, the Baranya-Evans finding could not be correct.

2) Saija, Prestipino, and Giaquinta in 2001 showed by simulations of both the Yukawa and the LJ liquids that the scaled radial distribution functions are identical for states where the excess entropy equals the two-particle entropy.\cite{sai01} This is consistent with the fact that these liquids are both strongly correlating, because scaled radial distribution function, excess entropy, and two-particle entropy are all isomorph invariants.

3) The basic idea of mode-coupling theory -- that statics determines the dynamics -- is consistent with the existence of isomorphs: For any strongly correlating liquid one expects that, if two state points have the same reduced pair-correlation functions, they are isomorphic. This means that they have the same (reduced) dynamics, so in this sense the pair correlation function ``determines'' the dynamics. From the isomorph perspective mode-coupling theory may be expected to work best for strongly correlating liquids.

4) Recently, Roland briefly reviewed characteristic times and their invariance to thermodynamic conditions from a general point of view, including also systems that are not liquids.\cite{rol08} He showed that quite different transitions in systems with slow relaxations -- onset of activated dynamics, dynamic crossover in viscous liquids and polymers, order-disorder transitions in liquid crystals, vitrification -- at varying temperature and pressure all take place at state points with the same value of the relaxation time. The conclusion is that\cite{rol08} ``the control parameter driving these transitions has the same functional dependence on $T$, $p$, and $V$ as the relaxation time.'' This follows from the existence of isomorphs, although we can only explain this observation for strongly correlating liquids and solids.

5) The concept of hidden scale invariance may have implications beyond liquid state theory. Thus recently Procaccia and collaborators studied by simulation plastic flow of amorphous solids in the athermal limit.\cite{ler09} For two-dimensional solids composed of multi-disperse particles with the interaction length taken from a Gaussian distribution they showed that stress-strain curves at different densities collapse to a master curve. This happens when stress is scaled by $\rho^\nu$, where $\nu$ is $5.87$ when the repulsive part of the potential can be fitted by the IPL term $r^{-10}$. The number $5.87$ is not far from the $5$ predicted by the two-dimensional exponent $\gamma=n/2$ for IPL potentials.

\section{Concluding remarks}

This paper introduces the concept of isomorphic curves in the state diagram of a strongly correlating liquid. The existence of isomorphs reflects the liquid's hidden scale invariance, and therefore the class of strongly correlating liquids is identical to the class of liquids with isomorphs (Appendix A). Isomorphs may be labelled by any of these invariants, for instance $\ccv$. The isomorph concept, in fact, may be justified by starting from the concept of an ``order parameter'' labelling curves in the state diagram of same physical properties: Suppose a liquid is described by a single ``order parameter'' in the sense that many of its properties are determined by the value of the ``order parameter''. In this case one may draw curves in the state diagram along which these properties are all invariant. If one asks how it is that several properties could possibly correlate in this way, the simplest answer is that for any two state points on a constant ``order parameter'' curve, the canonical probabilities are proportional for microscopic configurations that somehow correspond to each other -- and the simplest possibility is that the ``corresponding'' configurations are those that trivially scale into one another. This is nothing but the isomorph definition. Thus the isomorph concept may be arrived at via some extrapolation by postulating that single-order-parameter liquids exist and inquiring into their properties.

All isomorph invariants apply to IPL liquids. But as emphasized repeatedly, the converse is not true: Not all properties that are invariant for an IPL liquid along states obeying $\rho^{n/3}/T={\rm Const.\,}$  generalize to become isomorph invariants. The exceptions from approximate IPL behavior derive from the fact that the constant $C_{12}$ of Eq. (\ref{isodef}) generally differs from unity. Only IPL invariants that are independent of  the identity $C_{12}=1$, which characterizes IPL liquids, generalize to the class of strongly correlating liquids.

The present and previous papers on strongly correlating liquids show that this class of liquids is simpler than liquids in general. This is consistent with the general understanding among liquid-state specialists for many years, according to which non-associated liquids are generally simpler than associated ones. The virial / potential energy correlation coefficient $R$ provides a quantitative criterion for distinguishing simple liquids from the more complex -- and admittedly often more spectacular -- liquids that are not strongly correlating.

\acknowledgments 
In writing this paper we benefitted from discussions with from Hans Andersen, Austen Angell, Giulio Biroli, Simone Capaccioli, Daniele Coslovich, Pablo Debenedetti, Gregor Diezemann, Jack Douglas, Sharon Glotzer, Andreas Heuer, Valeria Molinero, Itamar Procaccia, Mike Roland, Ken Schweizer, Francesco Sciortino, Thomas Truskett, as well as in Roskilde with Morten Andersen, Ditte Gundermann, Claudio Maggi, Kristine Niss, S{\o}ren Toxv{\ae}rd, and Frank Vinther. This work was supported by a grant from the Danish National Research Foundation (DNRF) for funding the centre for viscous liquid dynamics ``Glass and Time.''

\appendix

\section{Equivalence of three conditions}

This Appendix proves that the following three conditions are equivalent for any liquid:

\begin{itemize}

\item (a) The liquid is strongly correlating.

\item (b) The liquid has isomorphs.

\item (c) The liquid has curves in the state diagram along which the reduced-coordinate constant potential energy hypersurface is invariant.

\end{itemize}

Only true IPL liquids obey these conditions rigorously (i.e., with 100\% correlation in (a)), and the equivalences may be stated more accurately as follows: A liquid is strongly correlating if and only if it has isomorphs to a good approximation, which happens if and only if the liquid has curves in the state diagram along which the reduced-coordinate constant potential energy hypersurface is almost invariant. It is instructive to prove all six implications, although logically $(a)\Rightarrow (b)\Rightarrow (c)\Rightarrow (a)$ would suffice.

Before proving the equivalences we note a few facts. If $\delta$ denotes the variation between two infinitesimally close microscopic configurations with same reduced coordinates, we first show that

\begin{equation}\label{du}
\delta U\,=\,
\left(d\ln\rho\right) W\,.
\end{equation}
For two infinitesimally close microscopic configurations with the same reduced coordinates ($\tilde{\bf r}_i\equiv\rho^{1/3} {\bf r}_i$) one has $0=\delta\tilde{\bf r}_i=(\rho^{-2/3}/3)(d \rho){\bf r}_i+\rho^{1/3}\delta{\bf r}_i$, i.e., $\delta{\bf r}_i=-(1/3)(d \ln \rho) {\bf r}_i$. The change of potential energy between the two configurations is given by $\delta U =\sum_i \delta{\bf r}_i\cdot{\bf\nabla}_{{\bf r}_i}U$, which via the definition of the virial, $W\equiv -1/3 \sum_i {\bf r}_i \cdot {\bf \nabla}_{{\bf r}_i}U$,  leads to Eq. (\ref{du}). We note further the following identity 

\begin{equation}\label{dut}
\delta\left(\frac{U}{T}\right)\,=\,
\frac{\left(d \ln\rho\right)W-\left(d \ln T\right)U}{T}\,.
\end{equation}
This follows by differentiation and subsequent use of Eq. (\ref{du}). Finally, note that the infinitesimal version of the isomorph condition Eq. (\ref{isodef}) is 

\begin{equation}\label{dutt}
\delta\left(\frac{U}{T}\right)\,=\,{\rm Const.}
\end{equation}
At any given state point $Q$, if the average potential energy is $\langle U\rangle_Q$, we define the constant potential energy hypersurface $\Omega$ as the subset of $R^{3N}$ given by

\begin{equation}\label{omegadef}
\Omega\,=\,
\{(\bR)\in R^{3N}\, |\,U(\bR)=\langle U\rangle_Q\}\,.
\end{equation}
The corresponding reduced-coordinate constant potential energy hypersurface $\tilde\Omega$ is given by

\begin{equation}\label{omegatildedef}
\tilde\Omega\,=\,
\{(\tR)\in R^{3N}\, |\,U(\rho^{-1/3} \tilde{\bf r}_1, ..., \rho^{-1/3} \tilde{\bf r}_N)=\langle U\rangle_Q\}\,.
\end{equation}

{\it Proof that $(a) \Leftrightarrow (b)$}: 
A strongly correlating liquid has (near) proportionality between virial and potential energy fluctuations, $\Delta W =\gamma \Delta U$. Thus at any given state point $W=\gamma U + C$ to a good approximation for the physically relevant microscopic configurations. If density and temperature are changed infinitesimally such that $d\ln T=\gamma\, d\ln\rho$, Eq. (\ref{dut}) implies that $\delta(U/T)={\rm Const.}$, which is the isomorph condition Eq. (\ref{dutt}). Suppose conversely that a liquid has isomorphs, and let $\Delta$ denote the difference between two arbitrary, physically relevant configurations at the state point in question. Then Eq. (\ref{dutt}) implies $\Delta \delta (U/T) = 0$ where $\delta$ (as usual) refers to changes from one configuration to another infinitesimally close by with the same reduced coordinates. Via Eq. (\ref{dut}) this implies $\left(d \ln\rho\right)\Delta W=\left(d \ln T\right)\Delta U$, i.e., the liquid is strongly correlating with $\gamma=d\ln T/d\ln\rho$. 

{\it Proof that $(a) \Leftrightarrow (c)$}: 
For a strongly correlating liquid $\Delta W \propto \Delta U$ for fluctuations between physically relevant configurations at any given state point. This implies that the hypersurfaces of constant virial and constant potential energy coincide. In particular, $W$ is constant on the state point's constant potential energy hypersurface $\Omega$. Accordingly, if density is changed infinitesimally, the change in potential energy between microscopic configurations with the same reduced coordinates is the same for all microscopic configurations on $\Omega$ (Eq. (\ref{du})). Thus a new hypersurface of constant potential energy is arrived at by slightly scaling $\Omega$; by adjusting temperature the new hypersurface is where the potential energy equals the average potential energy. Finally we note that the new and old hypersurfaces by construction have same reduced coordinates, thus the two state points have the same $\tilde\Omega$. Suppose conversely that two infinitesimally close state points have the same $\tilde\Omega$. All points on the two constant potential energy hypersurfaces differ by the same potential energy, which via Eq. (\ref{du}) implies that $W$ must be constant on each hypersurface. In other words, $W$ is constant on surfaces where $U$ is. This implies 100\% correlation between $W$ and $U$. For large systems the fluctuations are small, relatively, and a first order Taylor expansion of the $WU$ relationship leads to a linear relationship and 100\% correlation between $W$ and $U$.

{\it Proof that $(b) \Leftrightarrow (c)$}: 
Suppose a liquid has isomorphs. For a state point $Q$ on an  isomorph $I$ one concludes from Eq. (\ref{uinv}) that $\langle U\rangle_Q = k_BT \langle f_I\rangle +g(Q)$, where the (canonical) average $\langle f_I\rangle$ by Eq. (\ref{pinv}) is independent of $Q$. Consequently, $\tilde\Omega=\{(\tR)\in R^{3N}\,|\,k_BTf_I(\tR)+g(Q)=k_BT \langle f_I\rangle +g(Q)\} =\{(\tR)\in R^{3N}\,|\,\,f_I(\tR)=\langle f_I\rangle\}$ is invariant along the isomorph. Suppose conversely that two state points have the same $\tilde\Omega$. For these state points microcanonical averages are identical for all quantities that may be expressed as functions of the reduced coordinates. By the equivalence of the microcanonical and the canonical ensemble, for the two state points in question canonical ensemble averages are likewise identical for all quantities that are functions of the reduced coordinates. This can only be so if there is identity of the normalized canonical probability factors of any two physically relevant microscopic configurations of the two state points with the same reduced coordinates. This is another way of stating that the two state points are isomorphic.

\section{Properties of the density scaling exponent}

This appendix has two purposes: 1) To derive an optimization property of the density scaling exponent $\gamma$ of Eq. (\ref{gammaS}), and 2) to prove that any possible state-point dependence of $\gamma$ comes from a -- for strongly correlating liquids generally weak -- density dependence; more precisely it is shown that $\gamma=\gamma(\rho)$ in the isomorph approximation.

As mentioned in the main text there is no unique solution to the problem of finding the ``correct'' density scaling exponent $\gamma$, i.e., the exponent identifying isomorphs as the curves along which $\rho^\gamma/T={\rm Const.}$ First of all, the exponent must be expected to vary slightly with state point. But even at a given state point, there is no unique $\gamma$ in the sense that all isomorph invariants are mathematically unchanged for infinitesimal steps away from the state point in question obeying $\rho^\gamma/T={\rm Const.}$ This is because, except for IPL liquids, isomorph properties are approximate, so for instance the curves of constant excess entropy cannot be expected to be precisely the curves of constant isochoric specific heat (or constant reduced relaxation time, etc) -- this just applies to a good approximation.

We argued in Sec. IID that at any given state point there are three obvious gammas, the $\gamma_1$, $\gamma_2$, and $\gamma_3$ of Eq. (\ref{gammadef}). We recommend using $\gamma_1$ and used this in Sec. III when comparing isomorph predictions to computer simulations. This exponent is to be preferred because it -- among other things -- makes the excess entropy an exact invariant. It worked very well with the simulations of the KABLJ liquid, but as noted in the main paper $\gamma_2$ works almost equally well.

The excess entropy gamma $\gamma_1$, henceforth just denoted by $\gamma$, has an optimization property coming from answering the following question: At any given state point, suppose we change density by an infinitesimal amount. How much should temperature be changed to arrive at a new state point which is ``as isomorphic as possible'' with the original state point? To answer this we note that from Eq. (\ref{dutt}) one would require the quantity $F\equiv\delta(U/T)$ to be as constant as possible (where the symbol $\delta$ as previously refers to the difference between two infinitesimally close microscopic configurations with same reduced coordinates). This is obtained by minimizing $\langle(\Delta F)^2\rangle$. Since according to Eq.  (\ref{dut}) one has $F\propto (d \ln\rho) W-  (d \ln T) U$, for a given density change the quantity to be minimized by varying temperature is $\langle\big((d \ln\rho)\Delta W-(d \ln T)\Delta U\big)^2\rangle$, i.e.

\begin{equation}\label{minim}
(d \ln\rho)^2 \langle(\Delta W)^2\rangle + (d \ln T)^2 \langle(\Delta U)^2\rangle
-2(d \ln\rho)(d \ln T)\langle\Delta W \Delta U\rangle\,.
\end{equation}
Equating to zero the derivative of this expression with respect to $d \ln T$ leads to

\begin{equation}\label{sgammaigen}
\frac{d \ln T}{d \ln\rho}\,=\,
\frac{\langle\Delta W \Delta U\rangle}{\langle(\Delta U)^2\rangle}\,,
\end{equation}
which is the excess entropy gamma of Eq. (\ref{gammaS}). 

Suppose instead that one asks the complementary question: For a given an infinitesimal temperature change, what is the density change giving a new state point that is as ``isomorphic as possible'' with the original state point? In this case, minimizing Eq. (\ref{minim}) leads to the $\gamma_3$ of Eq. (\ref{gammadef}). Thus if density and temperature were equivalent variables, optimization arguments cannot determine which gamma to choose. In the NVT ensemble, however, temperature and density are not quite equivalent because, even though they are both externally controlled, the average kinetic energy fluctuates while the volume is strictly fixed.

The second property of the density scaling exponent to be proved is that if the $\gamma$ of Eq. (\ref{gammaS}) varies with state point, this variation can come only from a density dependence. More accurately, in the ``isomorph approximation'' where the curves of constant excess entropy and constant excess isochoric specific heat coincide, one has $\gamma=\gamma(\rho)$. Recalling that the excess pressure coefficient $\bv$ is defined by $\bv=(\partial (W/V)/\partial T)_V$ (Papers I and III) and denoting the excess isochoric specific heat per unit volume by $\cv$, the standard fluctuation expressions for these quantities (see, e.g., Appendix B of Paper I) implies that Eq. (\ref{gammaS}) may be written

\begin{equation}\label{gamma_frac}
\gamma\,=\,
\frac{\bv}{\cv}\,.
\end{equation}
If there is identity of the curves in the phase diagram of constant excess entropy $\sex$ and those of constant (extensive) excess isochoric specific heat $\ccv$ ($\ccv=V\cv$), the definition of the density scaling exponent $\gamma=(\partial\ln T/\partial\ln\rho)_{\sex}$ implies

\begin{equation}\label{g1}
\gamma\,=\,
-\left(\frac{\partial\ln T}{\partial\ln V}\right)_{\ccv}\,.
\end{equation}
From this we get via the mathematical identity $(\partial x/\partial y)_z(\partial y/\partial z)_x(\partial z/\partial x)_y=-1$

\begin{equation}\label{g2}
\left(\frac{\partial\ln\ccv}{\partial\ln T}\right)_V\,=\,
\frac{-1}{\gamma}\left(\frac{\partial \ln T}{\partial\ln V}\right)_{\ccv}
\left(\frac{\partial\ln\ccv}{\partial\ln T}\right)_V\,=\,
\frac{1}{\gamma}\left(\frac{\partial\ln\ccv}{\partial\ln V}\right)_T\,.
\end{equation}
Using $\ccv=T(\partial\sex/\partial T)_V$ we get via Eq. (\ref{gamma_frac})

\begin{equation}\label{g3}
\left(\frac{\partial\ln\ccv}{\partial\ln T}\right)_{V}\,=\,
\frac{\ccv}{V\bv}\left(\frac{\partial\ln\ccv}{\partial\ln V}\right)_T\,=\,
\frac{1}{\bv} \left(\frac{\partial\ccv}{\partial V}\right)_T\,=\,
\frac{T}{\bv} \frac{\partial^2\sex}{\partial V\partial T}\,.
\end{equation}
The Maxwell relation $(\partial\sex/\partial V)_T=(\partial (W/V)/\partial T)_V=\bv$ allows us to rewrite this as   

\begin{equation}\label{g4}
\left(\frac{\partial\ln\ccv}{\partial\ln T}\right)_{V}\,=\,
\frac{T}{\bv} \left(\frac{\partial\bv}{\partial T}\right)_V\,=\,
\left(\frac{\partial\ln\bv}{\partial\ln T}\right)_{V}\,.
\end{equation}
Because $d\ln\cv=d\ln\ccv$ at constant volume this implies that

\begin{equation}\label{g5}
\left(\frac{\partial\ln\bv}{\partial\ln \cv}\right)_{V}\,=\,
\left(\frac{\partial\ln\bv}{\partial\ln \ccv}\right)_{V}\,=1\,.
\end{equation}
Thus at constant volume $\bv$ is proportional to $\cv$, i.e., $\gamma$ is constant on isochores.

\section{Relating the correlation coefficient of Fig. 1 to the $WU$ correlation coefficient $R$ of Eq. (\ref{R})}

In order to distinguish from the $WU$ correlation coefficient $R$ of Eq. (\ref{R}), for the ``direct isomorph check'' of Fig. (\ref{fig:figure0}) we denote by $R_{DI}$ the correlation coefficient between potential energies of microscopic configurations with same reduced coordinates. Writing for a strongly correlating liquid for each microstate $\Delta W=\gamma \Delta U+\varepsilon$, where $\varepsilon$ is an ``error'' term uncorrelated with $\Delta U$, implies by squaring and averaging

\begin{equation}\label{W2}
\angleb{(\Delta W)^2} = \gamma^2 \angleb{(\Delta U)^2} + \angleb{\varepsilon^2}\,.
\end{equation}
Multiplying $\Delta W=\gamma\Delta U+\varepsilon$ by $\Delta U$, averaging and squaring gives

\begin{equation}
\angleb{\Delta W\Delta U}^2 = \gamma^2\angleb{(\Delta U)^2}^2.
\end{equation}
The left hand side of this is expressed in terms of the $WU$ correlation coefficient $R$ , giving 

\begin{equation}
\angleb{(\Delta W)^2} = \gamma^2\angleb{(\Delta U)^2}/R^2\,.
\end{equation}
Eliminating $\angleb{(\Delta W)^2}$ between this and Eq. (\ref{W2}) gives a relation between the correlation coefficient and the variance of the error term,

\begin{equation}\label{R2_var_eps}
\frac{1}{R^2} - 1 = \frac{\angleb{\varepsilon^2}}{\gamma^2\angleb{(\Delta U)^2}}\,.
\end{equation}

Next consider an infinitesimal rescaling of all microscopic configurations to a new density, calling the old and new potential energies of corresponding microscopic configurations $U^{(1)}$ and  $U^{(2)}$, respectively. If the relative density change is small, Eq. (\ref{du}) minus its average implies for each microscopic configuration

\begin{equation}
\Delta U^{(2)} = \Delta U^{(1)} + (d\ln\rho) \Delta W^{(1)},
\end{equation}
which, using $\Delta W=\gamma \Delta U+\varepsilon$, becomes 

\begin{equation}
\Delta U^{(2)} = (1+\gamma\, d \ln\rho)\Delta U^{(1)} + (d\ln\rho)\varepsilon\,.
\end{equation}
We can now calculate the correlation coefficient $R_{DI}$ between the old and new potential energies of microscopic configurations with the same reduced coordinates. In fact, since this equation is identical in structure to  $\Delta W=\gamma \Delta U+\varepsilon$, we can use the result of Eq.  (\ref{R2_var_eps}), replacing $\gamma$ with $1+\gamma\, d\ln\rho$ and $\varepsilon$ with $(d\ln\rho)\varepsilon$:

\begin{equation}
\frac{1}{R_{DI}^2} - 1 = \frac{\angleb{\varepsilon^2}(d \ln\rho)^2}{(1+\gamma\, d\ln\rho)^2 \angleb{(\Delta U)^2}}\,.
\end{equation}
Using Eq. (\ref{R2_var_eps}) to eliminate $\angleb{\varepsilon^2}/\angleb{(\Delta U)^2}$ finally implies to lowest order in $d \ln\rho$:

\begin{equation}\label{RRrel}
\frac{1}{R_{DI}^2} - 1 = \left(\gamma\, d\ln\rho \right)^2\left(\frac{1}{R^2} - 1\right)\,.
\end{equation}

Equation (\ref{RRrel}) applies to any liquid. If the density change goes to zero ($d\ln\rho\rightarrow 0$) and/or the liquid becomes 100\% correlating ($R\rightarrow 1$), there is perfect correlation between the new and old potential energies ($R_{DI}\rightarrow 1$), as expected.

\end{document}